\documentclass[a4paper,11pt]{article}
\pdfoutput=1 

\usepackage{fake_JNP} 
\usepackage{caption}

\usepackage[T1]{fontenc} 
\usepackage[all]{xy}
\newcommand{\vect}[1]{\boldsymbol{#1}}

\title{\boldmath Visual Information flow in Wilson-Cowan networks}

\author[a]{A. Gomez-Villa,}
\author[a]{M. Bertalm\'io,}
\author[b,1]{J. Malo\note{Corresponding author}}

\affiliation[a]{Dept. Inf. Comm. Tech. Universitat Pompeu Fabra, Barcelona, Spain}
\affiliation[b]{Image Processing Lab, Universitat de Val\`encia, Valencia, Spain}

\emailAdd{alexander.gomez@upf.edu}
\emailAdd{marcelo.bertalmio@upf.edu}
\emailAdd{jesus.malo@uv.es}

\abstract{\\
In this work we study the communication efficiency of a psychophysically-tuned cascade of Wilson-Cowan and Divisive Normalization layers
that simulate the retina-V1 pathway.
This is the first analysis of Wilson-Cowan networks in terms of multivariate total correlation.
The parameters of the cortical model have been derived through the relation between
the steady state of the Wilson-Cowan model and the Divisive Normalization model.

The communication efficiency has been analyzed in two ways:
First, we provide an analytical expression for the reduction of the total correlation among
the responses of a V1-like population after the application of the Wilson-Cowan interaction.
Second, we empirically study the efficiency with visual stimuli and statistical tools that were not available before:
(1)~we use a recent, radiometrically calibrated, set of natural scenes, and
(2)~we use a recent technique to estimate the multivariate total correlation in \emph{bits} from sets of
visual responses which only involves univariate operations, thus giving better estimates of the redundancy.

The theoretical and the empirical results
show that although this cascade of layers was not optimized for statistical independence in any way,
the redundancy between the responses gets substantially reduced along the neural pathway.
Specifically, we show that
(1)~the efficiency of a Wilson-Cowan network is similar to its equivalent Divisive Normalization model,
(2)~while initial layers (Von-Kries adaptation and Weber-like brightness) contribute to univariate
equalization, the bigger contributions to the reduction in total correlation come from the computation of
nonlinear local contrast and the application of local oriented filters, and
(3)~psychophysically-tuned models are more efficient (reduce more total correlation) in the more populated regions
of the luminance-contrast plane. These results are an alternative confirmation of the \emph{Efficient Coding Hypothesis}
for the Wilson-Cowan systems.
And from an applied perspective, they suggest that neural field models could be an option in image coding to perform image compression.
}

\begin{document}
\maketitle
\flushbottom

\section{Introduction}

The Wilson-Cowan equations \cite{Wilson73} and Divisive Normalization \cite{Carandini94} are alternative influential models of interaction between cortical neurons.
While the Divisive Normalization computation has been extensively studied from an information theoretic perspective \cite{Schwartz01,Malo06b,Malo10,Coen-Cagli12,Martinez18},
the efficiency of Wilson-Cowan networks has not been analyzed in such detail through accurate redundancy measures.

Such analysis is interesting not only because of the classical \emph{Efficient Coding Hypothesis} \cite{Barlow59,Barlow61,Barlow01}, which is the information-theoretic version of the \emph{free-energy principle} \cite{Friston12}, but also because of practical applications. Note that equivalent analysis in the Divisive Normalization case led to substantial improvements in image compression using brain inspired architectures \cite{Malo06a,ICLR17}.
In a low noise context \cite{Bell95,Bell97,Simoncelli01} optimal systems are those that minimize the redundancy among the components of the signal representation \cite{Bethge06,Malo10,Laparra15}.

In this paper we quantify how effective is each layer of the considered network in achieving the density factorization goal.
This information theoretic analysis is applied to a recent psychophysically-tuned Wilson-Cowan model \cite{Malo18,Malo19}, using
a recent radiometrically calibrated database of visual scenes \cite{Foster15,Foster16}, and using a
recent accurate statistical tool to measure redundancy \cite{Laparra11,ICLR19}.

The appropriate (multivariate) concept to describe the redundancy in vectors (or neural populations) is the \emph{total correlation} \cite{Watanabe60,Studeny98}.
Direct computation of total correlation from its definition is not straightforward because it involves multivariate probability density estimation.
That is why, in the past, different computationally convenient (bivariate) surrogates of the total correlation have been used. For instance,
the conditional probability between pairs of responses \cite{Buccigrossi98,Mumford99,Schwartz01,HyvarinenBubbles03,Malo06b}, the correlation between the energies of the responses \cite{HyvarinenTICA,HyvarinenISA,Hyvarinen09,Epifanio00,Epifanio03},
or the mutual information between pairs of responses \cite{Moulin01,Malo06a,Malo10}.

Now, by using a recent multivariate Gaussianization technique \cite{Laparra11,ICLR19}, total correlation can be estimated from (easy to compute) marginal equalizations,
so we can use the proper measure of communication efficiency in psychophysically-tuned networks.
This study, which measures total correlation on Wilson-Cowan responses for the first time,
generalizes previous analysis that used total correlation but were restricted to linear models to avoid estimation problems \cite{Bethge06},
or addressed simpler (just-one-layer) linear+nonlinear models with bivariate mutual information measures \cite{Malo10}.

\section{Models, materials and methods}

In this section we first describe the considered model for the retina-V1 pathway that consists of a cascade of equivalent Divisive Normalization and Wilson-Cowan modules.
Then, we present the visual stimuli over which the considered model will be applied.
And finally, we will review the statistical methods to assess the communication efficiency of the model.

\subsection{Model: A psychophysically-tuned Wilson-Cowan network}

\begin{figure}[!t]
	\centering
    \small
    \setlength{\tabcolsep}{2pt}
    \vspace{-1.50cm}
    \begin{tabular}{c}
    \hspace{-0cm}  \includegraphics[height=0.9\textheight]{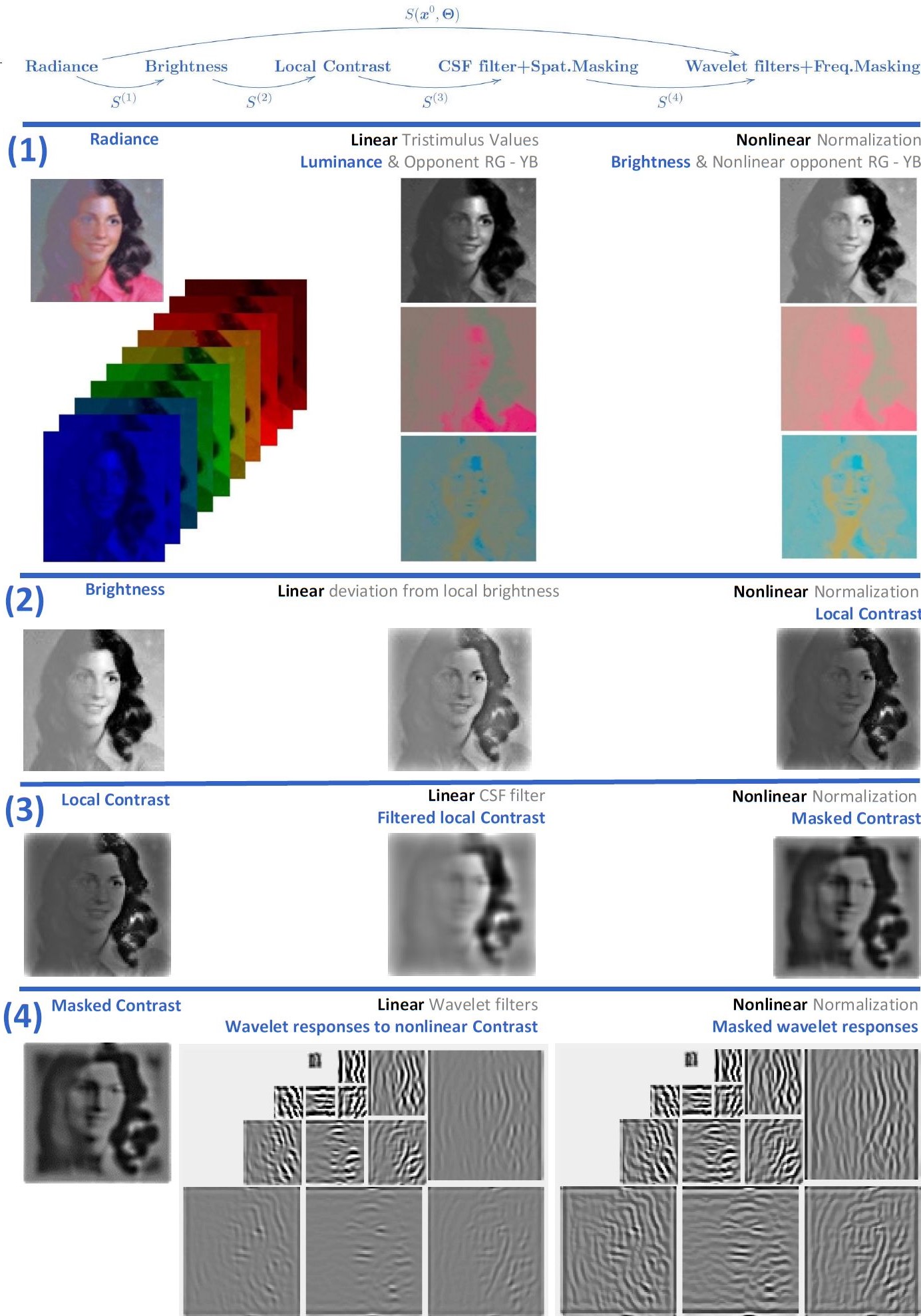} \\
    \vspace{-0.8cm}
    \hspace{-0cm}
   \end{tabular}
 \caption{\small{\textbf{Cascade of Linear+Nonlinear layers.} The network addresses in turn (1)~tristimulus-from-irradiance integrals and color opponency + Von-Kries adaptation together with Weber-like brightness and nonlinear opponent channels, (2)~local contrast (subtraction of local mean + division by local mean), (3)~CSF filter + masking in the spatial domain, and (4) Wavelet filters + masking in the wavelet domain. See details on the formulation of each Divisive Normalization layer in~\cite{Martinez18}. In this work we substitute the last nonlinearity by the equivalent Wilson-Cowan model.}}
             \label{model}
    \vspace{-0.3cm}
\end{figure}

In this work the theory is illustrated in the context of models of the retina-cortex pathway.
The considered framework follows the program suggested in \cite{Carandini12} and implemented in \cite{Martinez18}:
a cascade of four isomorphic \emph{linear+nonlinear} modules.
These modules address brightness, contrast, frequency filtered contrast masked in the spatial domain,
and orientation/scale masking. An example of the transforms of the input in such models is shown
in Fig. \ref{model}.

In this illustration the input is the spatial distribution of the \emph{spectral irradiance} at the retina. This input undergoes the following transforms: (1) The linear part of the first layer consist of three positive LMS spectral sensitivities and a linear recombination of the LMS values with positive/negative weights. This leads to three tristimulus values in each spatial location: one of them is proportional to the luminance, and the other two have opponent chromatic meaning (red-green and yellow-blue).
The linear tristimulus values are normalized by the corresponding values of the \emph{white} in the scene (Von-Kries adaptation), and then, the normalized opponent responses undergo parallel saturation transforms.
Perception of \emph{brightness} is mediated by an adaptive Weber-like nonlinearity applied to the luminance at each location. This nonlinearity enhances the response in the regions with small linear input (low luminance).
(2)~The linear part of the second layer computes the deviation of the brightness at each location from the local brightness. Then, this deviation is nonlinearly normalized by the local brightness
to give the local nonlinear contrast.
(3)~The responses to local contrast are convolved by center surround receptive fields (or filtered by the Contrast Sensitivity Function). Then the linearly filtered contrast is nonlinearly normalized by the local contrast. Again normalization increases the response in the regions with small input (low contrast).
(4)~After a linear wavelet transform modelling the response of simple cells in V1, each response is normalized by the activity of the neurons in the surround. Again, the activity relatively increases in the regions with low input. As shown in the marginal and joint PDFs below, the common effect of the nonlinear modules throughout the network is response equalization.

Divisive Normalization is the conventional model used to describe the nonlinearities in contrast perception psychophysics \cite{Foley94,Watson97}, but here we will also explore the equivalent Wilson-Cowan model in the last layer.


Below we introduce the notation of both interaction models (the Divisive Normalization and the Wilson-Cowan models),
and the relation between them so that we can infer psychophysically plausible parameters for the Wilson-Cowan model
from the parameters already tuned for Divisive Normalization.

\paragraph*{Modelling cortical interactions.}
In the case of the V1 cortex, we refer to the set of responses of a population of simple cells as the vector $\vect{r}$.
The considered models (Divisive Normalization and Wilson-Cowan) define a nonlinear mapping, $\mathcal{N}$, that transforms the input vector $\vect{r}$ (before the interaction among neurons) into the output vector $\vect{x}$ (after the interaction),
\vspace{-0.2cm}
\begin{equation}
  \xymatrixcolsep{2pc}
  \xymatrix{ \vect{r}  \,\,\,\, \ar@/^0.7pc/[r]^{\scalebox{0.85}{$\mathcal{N}$}} & \,\,\,\, \vect{x}
  }
  \label{global_response}
\end{equation}
In this setting, responses are called \emph{excitatory} or \emph{inhibitory}, depending on the corresponding \emph{sign} of the signal: $\vect{r} = \textrm{sign}(\vect{r}) |\vect{r}| $, and $\vect{x} = \textrm{sign}(\vect{x}) |\vect{x}|$.
The map $\mathcal{N}$ is an adaptive saturating transform, but it preserves the sign of the responses (i.e. $\textrm{sign}(\vect{x})=\textrm{sign}(\vect{r})$).
Therefore, the models care about cell activation (the modulus $|\cdot|$) but not about the excitatory or inhibitory nature of the sensors (the $\textrm{sign}(\cdot)=\pm$).

We will refer to as the \emph{energy} of the input responses to the vector $\vect{e} = |\vect{r}|^\gamma$, where this is an element-wise exponentiation of the amplitudes $|r_i|$.
Given the sign-preserving nature of the nonlinear mapping, for the sake of simplicity in notation, in the rest of the paper the variables $\vect{r}$ and $\vect{x}$ refer to the activations $|\vect{r}|$ and $|\vect{x}|$.

\vspace{-0.0cm}
\paragraph{The Divisive Normalization model.}
\emph{Forward transform:} The input-output transform in the Divisive Normalization is (in matrix notation \cite{Martinez18}),

\begin{equation}
    \vect{x} = \mathbb{D}_{\vect{k}} \cdot \mathbb{D}^{-1}_{\left( \vect{b} + \vect{H} \cdot \vect{e} \right)} \cdot \vect{e}
    \label{DN_B}
\end{equation}
\vspace{0.1cm}

\noindent where the output vector of nonlinear activations in V1, $\vect{x}$, depends on the energy of the input linear wavelet responses, $\vect{e}$, which are dimension-wise normalized by a sum of neighbor energies. Note that in this matrix notation, $\mathbb{D}_{\vect{v}}$, stands for a diagonal matrix with the vector, $\vect{v}$, in the diagonal.
The non-diagonal nature of the interaction kernel $\vect{H}$ in the denominator, $\vect{b} + \vect{H} \cdot \vect{e}$,
implies that the $i$-th element of the response may be attenuated if the activity of the neighbor sensors, $e_j$ with $j\neq i$, is high.
Each row of the kernel $\vect{H}$ describes how the energies of the neighbor simple cells attenuate the activity of each simple cell after the interaction.
The each element of the vectors $\vect{b}$ and $\vect{k}$ respectively determine the semisaturation and the dynamic range of the nonlinear response of each sensor.

\vspace{-0.0cm}
\emph{Inverse transform:} The relation between the two models is easier to obtain by identifying the corresponding
decoding transforms in both models. In the case of the Divisive Normalization, the analytical inverse is \cite{Malo06a,Martinez18},
\begin{equation}
      \vect{e} = \left( I - \mathbb{D}^{-1}_{\vect{k}}\cdot\mathbb{D}_{\vect{x}}\cdot \vect{H} \right)^{-1} \cdot \mathbb{D}_{\vect{b}} \cdot \mathbb{D}^{-1}_{\vect{k}} \cdot \vect{x}
      \label{invDN}
\end{equation}

\begin{figure}[b]
 \centering
 \small
 \includegraphics[height=5cm]{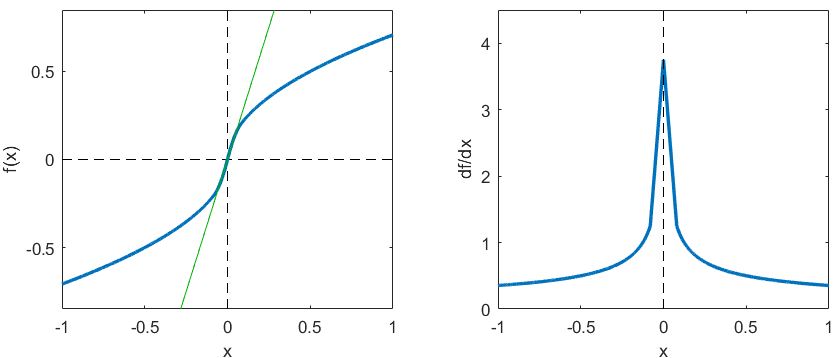}
 \caption{\small{\textbf{Saturation in the Wilson-Cowan model.}
\emph{Left:} Illustrative saturating function in blue and linear approximation around the origin in green. \emph{Right:} Derivative of the saturating function decreases with amplitude.}}
             \label{f_x}
\end{figure}

\vspace{-0.0cm}
\paragraph{The Wilson-Cowan model.}
\emph{Dynamical system:} In the Wilson-Cowan model the variation of the activation vector, $\vect{\dot{x}}$, increases with the energy of the input, $\vect{e}$, but, for each sensor, this variation is also moderated by its own activity and by a linear combination
of the activities of the neighbor sensors,
\vspace{-0.0cm}
\begin{equation}
      \vect{\dot{x}} = \vect{e} - \mathbb{D}_{\vect{\alpha}} \cdot \vect{x} - \vect{W} \cdot f(\vect{x})
      \label{EqWC}
\end{equation}
where $\vect{W}$ is the matrix that describes the damping factor between sensors, and $f(\vect{x})$ is a dimension-wise saturating nonlinearity (see Fig. \ref{f_x}).
Note that in Eq.~\ref{EqWC} both the inhibitory and the excitatory responses are considered just as negative and positive components of the same vector.
Therefore, following \cite{Bressloff03}, the two equations in the traditional Wilson-Cowan formulation are represented here by a single expression.

\emph{Steady state and inverse.}
The stationary solution of the above differential equation, $\vect{\dot{x}} =0$ in Eq.~\ref{EqWC}, leads to the following
decoding (input-from-output) relation:
\begin{equation}
      \vect{e} = \mathbb{D}_{\vect{\alpha}} \cdot \vect{x} + \vect{W} \cdot f(\vect{x})
      \label{invWC}
\end{equation}

The identification of the decoding equations in both models, Eq.~\ref{invDN} and Eq.~\ref{invWC}, is the key to obtain simple analytical relations between their parameters.

\paragraph{Equivalence of models.} The analytical relation between the steady state of the Wilson-Cowan model and the Divisive Normalization was originally proposed at the Conference celebrating the 50th anniversary of Prof. Cowan at the University of Chicago \cite{Malo18}, but detailed demonstration and discussion of the properties are given in \cite{Malo19}.

In summary, just to get a simpler analytical relation between the parameters, in \cite{Malo19}, an assumption was done in each model:
(1) a first order approximation of the nonlinear saturation of the Wilson-Cowan system (green line in Fig. \ref{f_x}), and (2)
a first order expansion of the inverse in decoding the Divisive Normalization (in Eq.~\ref{invDN}).
In this way, it is easy to see that the parameters of both models introduced above are related as:
\begin{eqnarray}
      \vect{\alpha} &=& \frac{\vect{b}}{\vect{k}} \nonumber \\
      \vect{W} &=& \mathbb{D}_{\left(\frac{\vect{x}}{\vect{k}}\right)} \cdot \vect{H} \cdot \mathbb{D}_{\left(\frac{\vect{k}}{\vect{b}} \odot \frac{df}{dx}\right)}^{-1}
      \label{relation_W_H}
\end{eqnarray}

\noindent where the $\odot$ stands for the dimension-wise Hadamard product, and the divisions are also Hadamard quotients.
This expression allows us to obtain the interaction kernel and the attenuation of a Wilson-Cowan model which leads to a
steady state compatible with the Divisive Normalization response.

\paragraph{The considered Wilson-Cowan model: parameters and performance.}
In this work we take an architecture as the one considered in Fig.~\ref{model}, in which different Divisive Normalization layers were obtained through different experimental methods. For instance, Maximum Differentiation psychophysics was used to get the 2nd and 3rd layers~\cite{Malo15}, the 1st layer was obtained by fitting human opinion in subjective image distortion~\cite{Martinez18}, and the last layer was tuned to reproduce contrast response curves~\cite{Martinez19}.

\begin{figure}[t!]
 \small
 \hspace{-0.8cm} \includegraphics[width=1.05\textwidth]{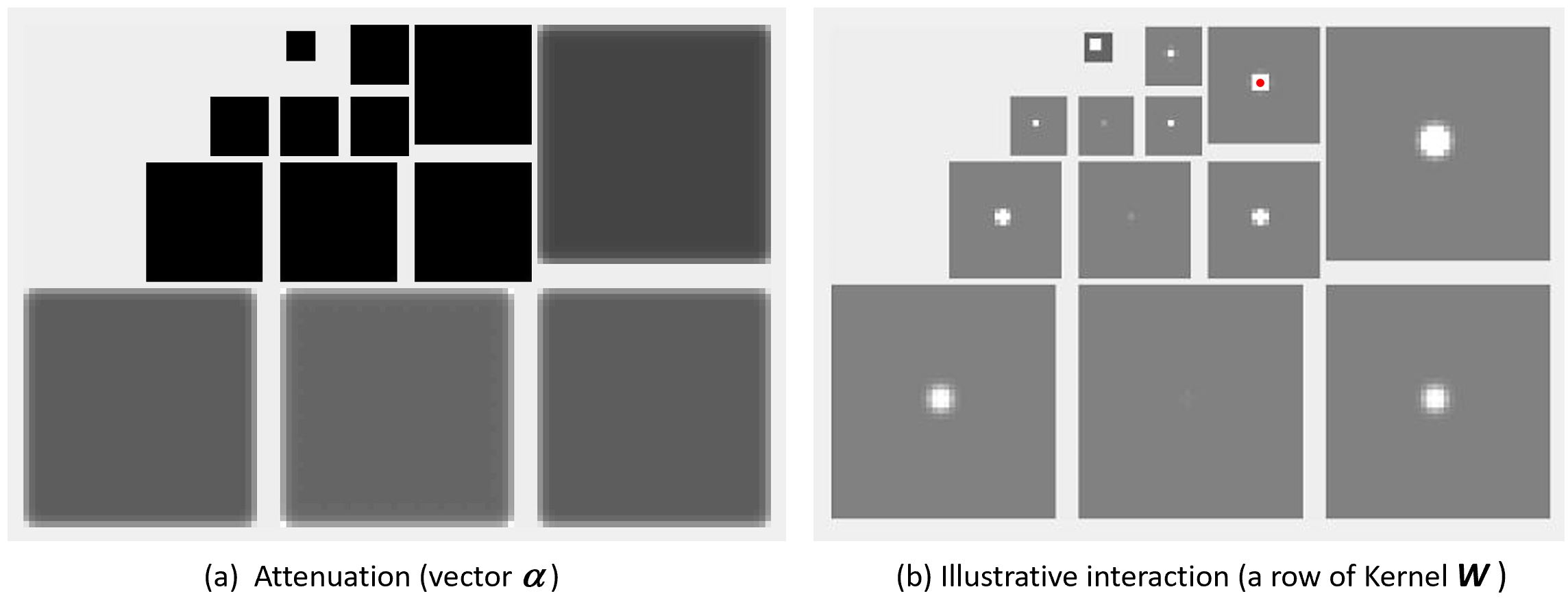}
 \vspace{-0.2cm}
 \caption{\small{\textbf{Parameters in the Wilson-Cowan cortical layer derived from the Divisive Normalization cortical layer.}
 \emph{Left:} The wavelet diagram represents the auto-attenuation values,~$\alpha_i$, for each sensor of the V1-like population. Gray values are linearly scaled to represent attenuation values from the minimum
 (black, for low frequencies) to the maximum (white, for high frequencies).
 \emph{Right:}~Illustrative interaction kernel for a specific coefficient (the one highlighted in red).
 Here lighter gray represents stronger interaction between the considered sensor and the rest of the sensors in the wavelet-like structure. Darker gray represents zero interaction.
 This diagram represents the corresponding row of $\vect{W}$ arranged as a wavelet vector.
 In both cases, $\vect{\alpha}$ and $\vect{W}$, were computed from the average response, $\vect{\hat{x}}$, in a large image database using the equivalence relation in Eq. \ref{relation_W_H}. The width of the Gaussian kernel in space is 0.08 deg, in scale is 1.1 octave, and in orientation is $\pi/6$ rad, and the values of the attenuation vector are in the range $[3\times10^3,1\times10^5]$.}}
\label{parameters_WC}
\end{figure}

In this psychophysically-tuned network, the nonlinearity in the cortical layer (the 4th layer consisting of a linear wavelet transform followed by Divisive Normalization) was substituted by the equivalent Wilson-Cowan interaction.
Following \cite{Wilson73,Faugueras09} we assumed a Gaussian interaction between the sensors tuned to different locations, scales and orientations, and it makes sense to have different attenuation coefficients per subband. However, instead of performing additional psychophysics to determine these parameters,
here we took the Divisive Normalization layer optimized in \cite{Martinez18,Martinez19}, and we applied the relation proposed in \cite{Malo18,Malo19}, i.e. Eq.~\ref{relation_W_H}.
The resulting kernel and attenuation for the Wilson-Cowan interaction are shown in Fig.~\ref{parameters_WC}.

\begin{figure}[t]
 \centering
 \vspace{0cm}
 \begin{tabular}{cc}
 \hspace{0cm}\includegraphics[height=0.44\linewidth]{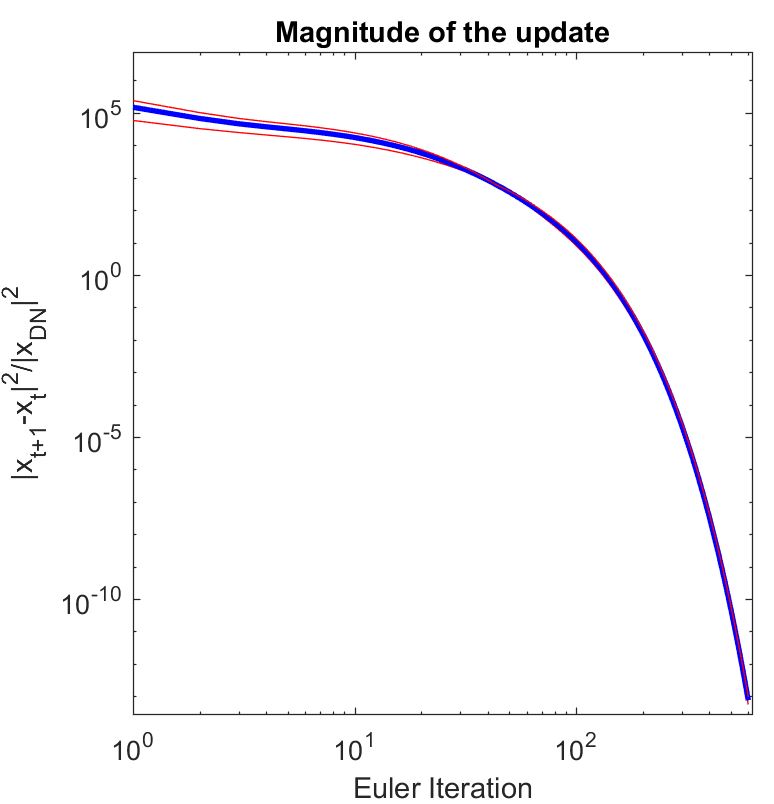} & \includegraphics[height=0.44\linewidth]{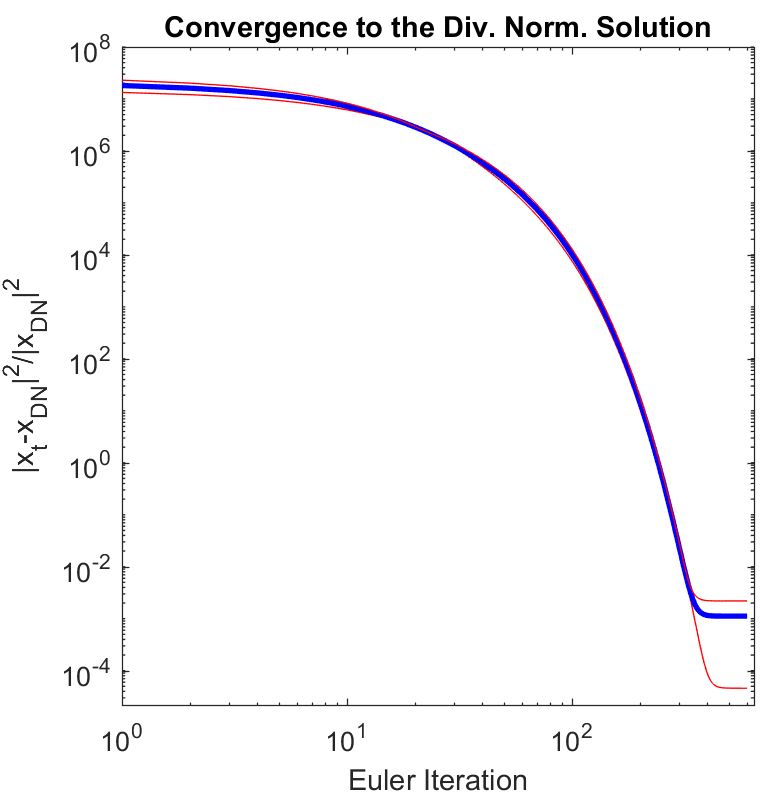}\\[-0.2cm]
 \end{tabular}
 \caption{\small{\textbf{Convergence of the Wilson-Cowan model to the Divisive Normalization solution.} \emph{Left}: evolution of the relative energy of the update of the solution along the integration.  \emph{Right}: evolution of the relative energy of the difference between the Wilson-Cowan solution and the Divisive Normalization response along the integration. The curves in blue are the average of the update and difference over 35 natural images of the Van Hateren database \cite{VanHateren98}, and the intervals in red represent 3 standard deviations below and above the mean.}}
 \label{convergence}
 \vspace{-0.3cm}
\end{figure}

Using these parameters, simple Euler integration from an initial output given by the input energy converges to the Divisive Normalization solution, see Fig.~\ref{convergence}.
A more detailed discussion on the Divisive Normalization as a stable node of this Wilson-Cowan system (theoretical stability, phase diagrams) is given in \cite{Malo19}, together with evidences of their perceptual equivalence beyond the mathematical equivalence.

Perceptual evidences in \cite{Malo19} were only focused on the \emph{visual consistence} between the original psychophysically-tuned kernel, $\vect{H}$, and the signal-dependent kernel obtained from Eq.~\ref{relation_W_H}, but \emph{always} within the Divisive Normalization context.
Specifically, it was shown that Divisive Normalization with these two different kernels has similar contrast response curves, and achieves similar correlation with human opinion in subjective image quality.

In this introductory section we present a new (more direct) evidence of the perceptual plausibility of the Wilson-Cowan model derived from the psychophysically-tuned Divisive Normalization.
In the image quality context, here we explicitly compute the response of the dynamic Wilson-Cowan model (integrating the equation until the steady state is achieved) for the original and the distorted images.
And then we check if the proposed Wilson-Cowan representation is more perceptually Euclidean than the input representation (the response after the three initial layers plus the linear wavelet transform).

\begin{figure}[t]
 \centering
 \small
 \hspace{0cm} \includegraphics[width=0.8\textwidth]{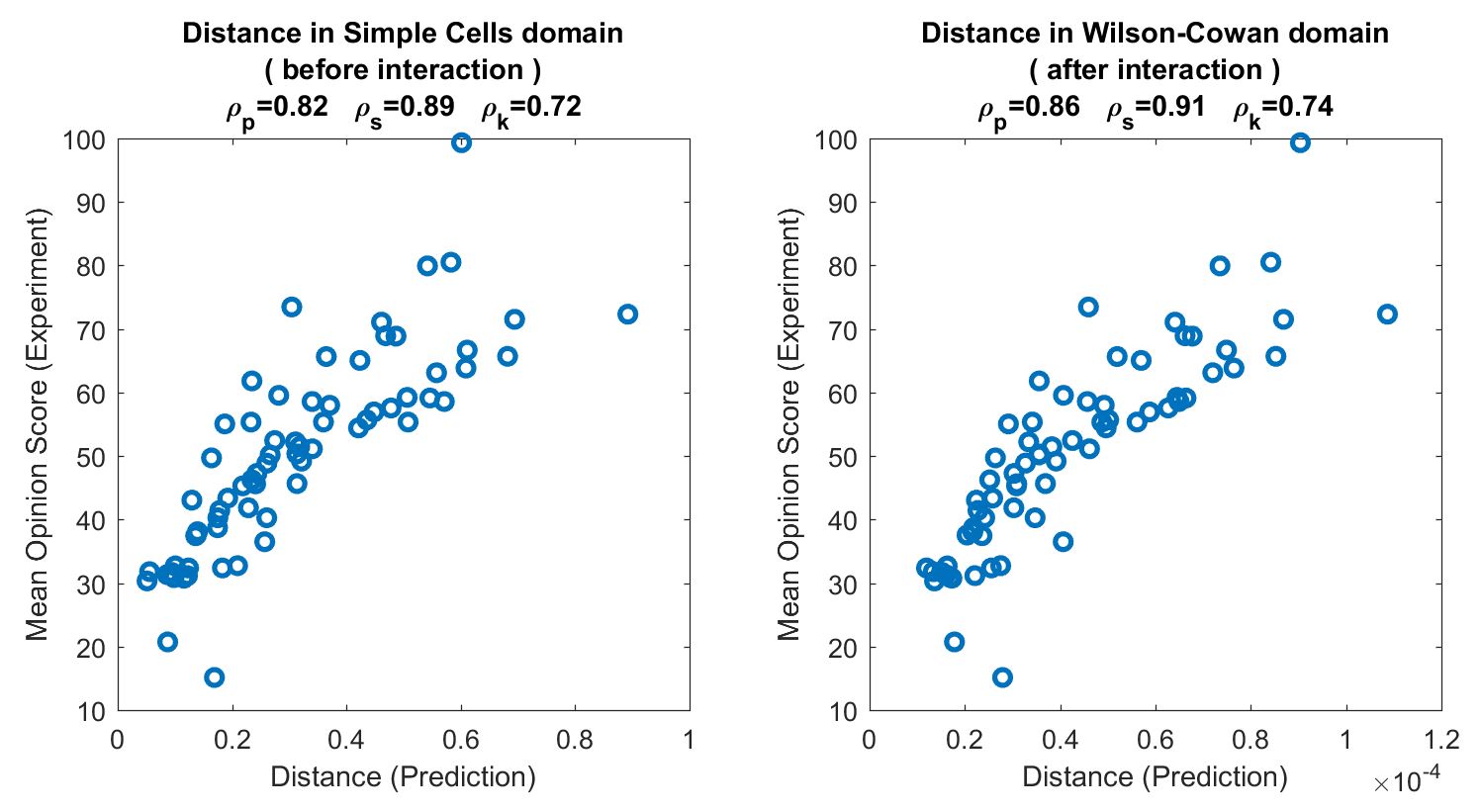}
 \caption{\small{\textbf{Wilson-Cowan improves the description of image distortion.}
 \emph{Left:} Alignment between subjective opinion and distance in the input representation.
 \emph{Right:}~Alignment between subjective opinion and distance in after the proposed Wilson-Cowan interaction. Stimuli and ground truth were taken from the subjectively rated image quality database TID \cite{ponomarenko08}.}}
\label{correlation}
\end{figure}

In Fig.~\ref{correlation} we compare the correlation between the experimental subjective visibility of distortions and the Euclidean distance between responses computed in the wavelet domain,~$\vect{r}$, and after the convergence of the Wilson-Cowan network, i.e. in the representation $\vect{x}$.
In our implementation (restricted to patches subtending 0.63 degrees of visual angle), this
illustration including $68$ distortions applied to one full-size image of the database, implied the integration of the Wilson-Cowan equation in 8970 image regions.
The result shows that the considered Wilson-Cowan model does improve the description of the perceived distortion in naturalistic environments with regard to the previous layer of the model.

In summary, the proposed Wilson-Cowan model, with parameters obtained according to Eq.~\ref{relation_W_H}, has the proper mathematical and perceptual behavior:
(a) it converges to the Divisive Normalization solution (Fig.~\ref{convergence}),
and (b) it describes the visibility of distortions better than the image representation given
by the linear simple cells (Fig.~\ref{correlation}).

\subsection{Material: radiometrically calibrated stimuli and color adaptation}

\paragraph{Scenes.} In this work we use stimuli from the databases of Foster and Nascimento \cite{Foster15,Foster16} because (a)~the scenes consist of radiometrically calibrated spectra in each pixel, and (b)~they also include information about the illumination through gray spheres of controlled reflectance. While the spectral nature of the scenes is useful to simulate the perception process from the initial integration over wavelengths \cite{Stiles82,Stockman11}, the information about the illumination allows a straightforward implementation
of Von-Kries adaptation with no extra gray-world assumptions \cite{Fairchild13}.

Alternative possibilities to perform Von-Kries adaptation in sensible LMS spaces include the Barcelona database \cite{Parraga09}, in which CIE XYZ images also include gray spheres; and the IPL database \cite{Laparra12,Gutmann14}, in which CIE XYZ scenes were illuminated using standard CIE~D65 and CIE~A spectra.

\begin{figure}[t]
 \small
 \hspace{-1.5cm} \includegraphics[width=1.2\textwidth]{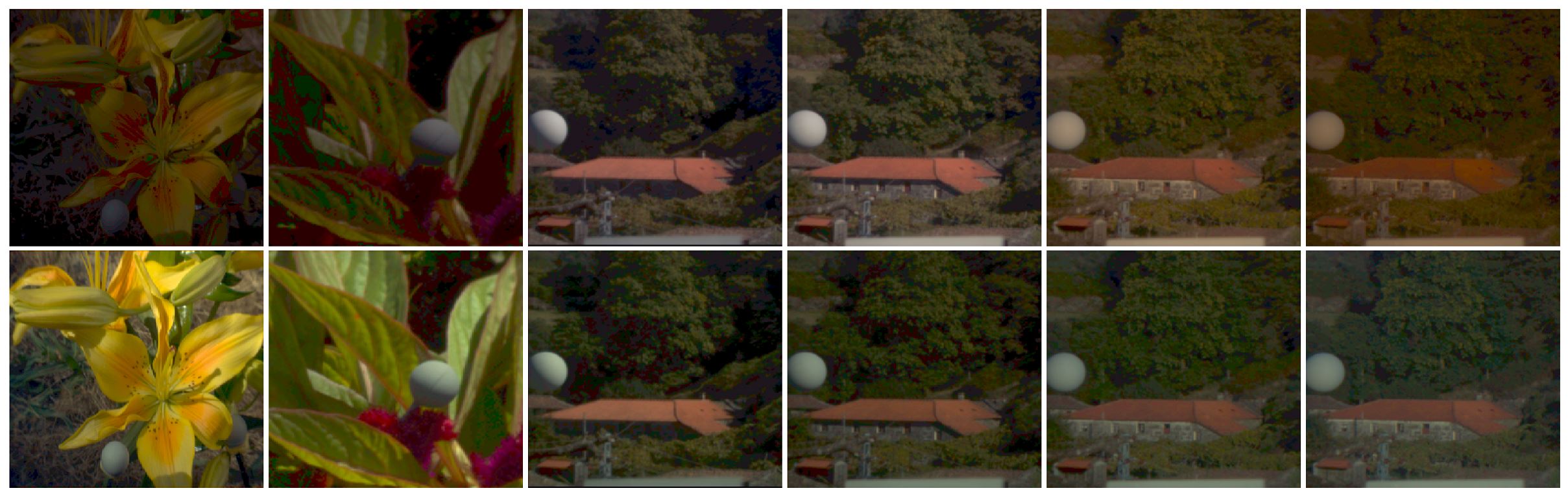}
 \caption{\small{\textbf{Examples of Von-Kries adaptation.}
 \emph{Top:} Representative scenes from \cite{Foster15,Foster16} under the original illumination. Spectral integration and \emph{tristimulus to digital counts} rendering was done with Colorlab \cite{Colorlab}.
 Note that (in the common illumination setting of the figure) the appearance of the reference gray is quite variable in the different scenes (different brightness, hue, and even saturation).
 \emph{Right:}~Same scenes after Von-Kries adaptation. All the gray references are normalized so their appearance is more stable and also the gamut of colors in the scenes.}}
\label{vonkries}
\end{figure}

\paragraph{Standard chromatic adaptation transforms.} We applied the Stockman and Sharpe LMS fundamentals \cite{Stockman00} to 5700 spectral image patches of size $40\times40\times33$ (including 33 wavelegths in the [380,700] nm range). In this way we got 5700 tristimulus image patches of size $40\times40\times3$.
We assumed that these patches subtend 0.625 degrees of visual angle. This implies assuming certain observation distance and spatial sampling frequency (in our case 64~cycles/deg), which is important to apply spatially calibrated models.
Each of these patches was associated to a specific scene, and the \emph{white point} of each scene was computed from the average LMS tristimulus values in the gray spheres of the scene (segmented by hand).
Each patch in the original LMS representation was Von-Kries normalized~\cite{Fairchild13} by the tristimulus values of the corresponding \emph{white point} leading to 5700 Von-Kries-adapted LMS images of size $40\times40\times3$.
Fig.~\ref{vonkries} shows the transform of different scenes to a common (Von-Kries adapted) canonical chromatic representation.

\paragraph{Information-theoretic considerations on the color manifold.}
The above chromatic adaptation stage reduces the variability and simplifies the structure of the color manifold in the LMS space because tristimulus measurements corresponding to objects seen under different illuminations are aligned and represented in a common scale~\cite{Webster95,Webster97,Laparra12,Laparra15}.
As a result, Von-Kries necessarily reduces the entropy of the color manifold.
Additionally, the considered linear opponent transform \cite{Ingling77} is similar
to PCA \cite{Buchsbaum83}, so it strongly reduces the redundancy between the color components.
Even though the actual nonlinearities applied to the linear opponent spaces involve the three chromatic channels \cite{Fairchild13,Laparra12}, here
we took a simplified (dimension-wise) approximation \cite{Stockman11}: we applied saturating nonlinearities only depending on the average tristimulus value in the scene in each separated channel to get the brightness and the corresponding nonlinear RG and YB values.
These additional nonlinearities should not modify the redundancy shared among the linear opponent tristimulus values because the total correlation is invariant under dimension-wise transforms \cite{Cover06,Studeny98}.

Despite the deep impact of these transforms on the redundancy among the components of the spectra and the components of the colors, in this work we restrict ourselves to the effects of the transforms on the spatial information since the Wilson-Cowan interaction considered above acts on spatial features. Further work may use the data available in the supplementary material and the statistical methods reviewed below to compare the efficiency of the (relatively simple) chromatic transforms used here with regard to more sophisticated color appearance models \cite{Fairchild13}, or other color representations specifically designed for information maximization or error minimization \cite{MacLeod03b,Laparra12,Laparra15}.

\subsection{Method: measuring Total Correlation with Gaussianization transforms}

Information transference from stimuli to response is maximized if the components of the inner representation are statistically independent \cite{Bell95,Bell97,Oja01}.
Therefore, the appropriate description of the communication efficiency of a perception system
consist of tracking the amount of information shared by the different components of the signal along the neural pathway.
This redundancy (or shared information) is the \emph{total correlation} \cite{Watanabe60} or \emph{multi-information} \cite{Studeny98}. The total correlation, $T$, is the Kullback-Leibler divergence between the joint PDF and the product of its marginal PDFs. Unfortunately, direct computation of $T$ from its definition is not straightforward because it involves multivariate PDF estimation in spaces with a huge number of dimensions.

The problems for direct computation of $T$ imply that the quantification of the efficiency of image representations, $\vect{x} = \mathcal{N}(\vect{r})$, is done using the variations $\Delta T$ from the input, $\vect{r}$, to the output, $\vect{x}$ \cite{Studeny98}:
\begin{equation}
      \Delta T = T(\vect{r}) - T(\vect{x}) = \sum_{i=1}^{d} (h(r_i) - h(x_i)) + \mathbb{E}\left( log |\nabla_{\!\vect{r}} \, \mathcal{N}(\vect{r})| \right)
      \label{eq_T_studeny}
\end{equation}
where the term with the sum over the dimensions is easy to compute because the marginal entropies, $h(\cdot)$, only depend on univariate PDF estimations, but the expected value of the Jacobian of the transform is required.
This second term complicates the estimation and, as a result, sometimes the analysis is restricted to linear transforms \cite{Bethge06}, where this second term is just a constant;
or surrogates of total correlation have to be used, as for instance multiple measures of mutual information between pairs of responses \cite{Malo06a,Malo06b,Malo10}, which only involve bivariate PDF estimations.

In this work we solve the above problems by using a novel estimator of $T$ which only relies on univariate estimations: the Rotation-Based Iterative Gaussianization (RBIG) \cite{ICLR19}.
The RBIG is a cascade of nonlinear+linear layers, each one made of (easy) marginal Gaussianizations followed by an (easy) rotation. This invertible architecture is able to transform any input PDF into a zero-mean unit-covariance multivariate Gaussian even if the chosen rotations are random \cite{Laparra11}.
This ability to completely remove the structure of any PDF is useful to estimate $T$ of arbitrary vectors $\vect{x}$: as the redundancy of a Gaussianized signal is zero, $T(\vect{x})$ corresponds to the sum of the individual variations, $\Delta T_l$, that take place along the layers of RBIG while Gaussianizing $\vect{x}$. Interestingly, the individual variation in each RBIG layer only depends on (easy to compute) marginal negentropies \cite{Laparra11}:
\begin{equation}
      T^{RBIG}(\vect{x}) = \sum_{l=1}^{L} \Delta T_l = \sum_{l=1}^{L} J_m(\vect{x}^{(l)})
      \label{eq_T_rbig}
\end{equation}
because the marginal negentropies, $J_m( \vect{x}^{(l)} )$, are just the sum of divergences between the marginal PDFs of the signal that is being Gaussianized at each layer (the vectors $\vect{x}^{(l)}$) and a univariate Gaussian.

In the \emph{Results} section the theoretical predictions on efficiency (obtained from the analytical Jacobian of a Wilson-Cowan system plugged into Eq.~\ref{eq_T_studeny}) are empirically confirmed by the computationally-convenient RBIG total correlation estimate, Eq.~\ref{eq_T_rbig}.

\section{Results: Redundancy reduction via the Wilson-Cowan interaction}

Here we study the evolution of the statistical dependence between the responses along the psychophysically-tuned neural pathway described in \emph{Models, Materials and Methods}.

First, we present an analytical result for the reduction of total correlation due to a Wilson-Cowan interaction. We check the validity of this theoretical result with an illustrative reduced-scale system. This reduced-scale example is useful not only to confirm the theory but also to illustrate the accuracy of the RBIG estimates of total correlation.

Then, we empirically analyze the behavior of the full-scale model in different ways: (1) by analyzing the shape of the marginal PDFs for different kinds of sensors at different layers of the network, (2) by computing the \emph{mutual information}, $I$, between the responses of multiple pairs of sensors at different layers, and finally, (3) by computing the (more appropriate) \emph{total correlation}, $T$, among the responses at the different layers.

\subsection{Theoretical analysis}

Here, we first present an analytical expression for the most interesting (multivariate) term in
the reduction of total correlation for the specific case of the Wilson-Cowan interaction.
Afterwards, we exhaustively check the validity of this expression in a reduced-scale example, and
we compare the efficiency of the Wilson-Cowan system with the efficiency of the equivalent
Divisive Normalization system.

\subsubsection{Expression for Total Correlation in Wilson-Cowan systems.}
The problem in estimating the variation of total correlation under arbitrary transforms (Eq. \ref{eq_T_studeny}) is the term depending on the Jacobian.
The determinant $|\nabla_{\vect{r}} \mathcal{N}(\vect{r})|$ represents the compression or expansion of the response space at the point $\vect{r}$.
Therefore, the last term is the \emph{average} variation of the volume of the response space over the PDF of natural signals.
As shown in the experiments below, this term determines the basic trends of the redundancy reduction.

In this section we give an analytical expression for this Jacobian-dependent term for a Wilson-Cowan interaction. We use this expression to predict how the efficiency of the system is going to be for natural images of different luminance/contrast.

%

Note that in the steady state of the Wilson-Cowan system there is a straightforward expression for the inverse (eq.~\ref{invWC}).
Taking into account that $\nabla_{\vect{r}} \mathcal{N}(\vect{r}_0) = (\nabla_{\vect{x}} \mathcal{N}^{-1}( \vect{x}_0 ))^{-1}$, $\forall \vect{r}_0$, where $\vect{x}_0 = \mathcal{N}(\vect{r}_0)$, the critical term of Eq.~\ref{eq_T_studeny} is:
\begin{equation}
       \mathbb{E}\left( log | \nabla_{\vect{r}} \mathcal{N} | \right) =
       log(\gamma) - \mathbb{E}\left( log( |\mathbb{D}_{\vect{\alpha}} + \vect{W} \cdot \mathbb{D}_{\frac{df}{dx}} | ) \right) - (\gamma^{-1}-1) \sum_{i=1}^d \mathbb{E}\left( log( \alpha_i x_i + v_i) \right)
      \label{analitic}
\end{equation}
where $\mathbb{D}_{\frac{df}{dx}}$ is a diagonal matrix with the derivative of the sigmoidal functions in each element of the diagonal, and $\vect{v} = \vect{W} \cdot f(\vect{x})$.

Eq. \ref{analitic} is interesting because one can infer the redundancy reduction as a function of relevant visual features such as luminance and contrast.
First, note that the activity of cortical sensors tuned to DC (flat patterns) increase with luminance and the activity of those tuned to AC (textured patterns) increase with contrast. Then, note that the first term in Eq.~\ref{analitic} is a constant, but the derivative in the second term decreases with contrast (see Fig.~\ref{f_x}), so this \emph{negative} term subtracts less for bigger contrasts.
On the contrary, the last term (also negative because typically $\gamma^{-1} > 1$) quickly increases with luminance and contrast, note that both components of this term increase with $\vect{x}$.

The above considerations suggest that, for constant contribution of the marginal entropies (which is the case, as shown below), the efficiency of
a Wilson-Cowan network should be bigger in the low-luminance / low-contrast region of the image space.
This trend suggested by the analysis of the terms in Eq.~\ref{analitic} is interesting because natural images typically have low luminance and low contrast \cite{Mumford99,Simoncelli98,malo2000role,Simoncelli01}.

\subsubsection{Experiment in a reduced-scale system.}
Here we consider a simplified example with the basic elements of the considered network but in a reduced-scale scenario: 3-pixel images.
In this situation the Jacobian matrices are small so the theory can be visualized and systematically checked.
The structure of this reduced-scale perception system is as follows:
first, 3-pixel luminance images are transformed into brightness through a pixel-wise (dimension-wise) Weber-like saturation \cite{Fairchild13}.
Then, 3 Fourier-like analyzers extract the DC, the low-frequency, and the high-frequency components of the brightness simulating frequency-selective cortical filters \cite{Watson87b}.
This frequency representation is weighted by a low-pass transfer function that simulates contrast sensitivity \cite{Campbell68}.
Finally, this set of responses interact either according to a static Divisive Normalization transform, or dynamically through the equivalent Wilson-Cowan equation. The specific parameters of this reduced-scale system are given in Appendix \ref{reduced-scale}.

\begin{figure}[t]
 \hspace{-2cm} \includegraphics[width=1.25\textwidth]{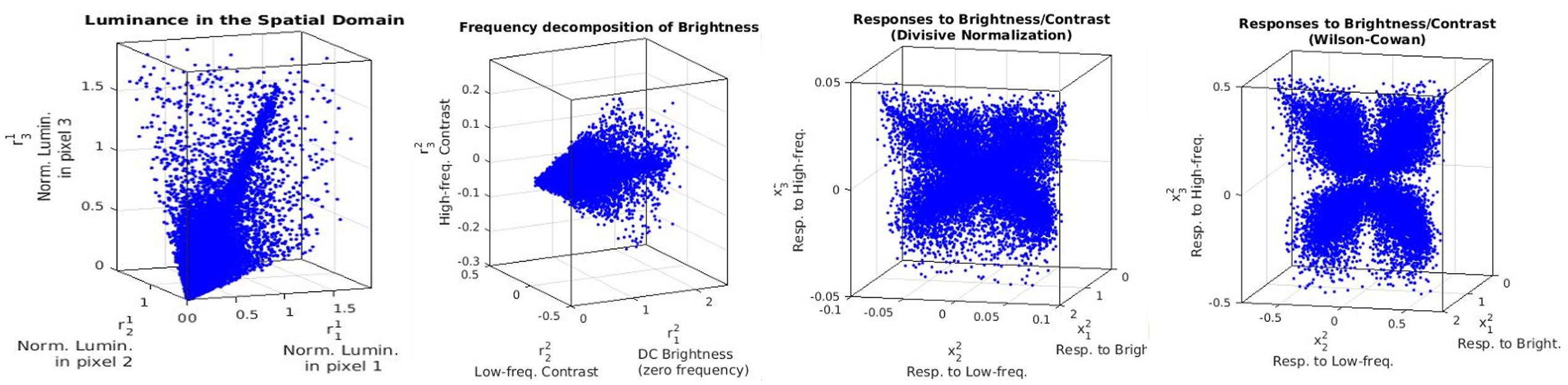}
 \caption{\small{\textbf{Natural images through the reduced-scale system.} The scatter plots ($2\times10^3$ samples randomly chosen from a set of $5\times10^6$ samples used in the experiments) display the changes on the PDF of the responses of the system to 3-pixel natural images at different layers of the network. \emph{From left to right:} (1) linear responses to luminance at the retina, (2) responses of frequency analyzers applied to brightness (saturated luminance), and (3) two versions of the nonlinear interaction at the cortex: the Divisive Normalization and the Wilson-Cowan responses. Absolute luminance (in cd/m2) has been normalized by the 95th percentile luminance value.}}
\label{manifolds}
\end{figure}

Figure \ref{manifolds} shows how this system transforms the manifold of 3-pixel luminance images of the considered database.
First, note how the frequency sensors (second scatter plot) look for the axes of symmetry of the manifold of natural images in the input domain, similarly to PCA \cite{Hancock92}.
However, the luminance-to-brightness saturating transform (scatter plot not shown) expands the low-luminance region and compresses the high-luminance region so the tail of brighter stimuli (high values in $r^2_1$ in the second scatter plot) is shorter. This will be discussed in further detail in analyzing the behavior of the full-scale model. Finally, note how the two interaction schemes considered here lead to a sort of factorization of the PDF in four separate blobs in the AC frequency components. This effect is also discussed in the next section with the marginal PDFs of the full-scale model in the wavelet domain.

The deformation of the manifolds shown in Fig.~\ref{manifolds} is an interesting illustration of the \emph{multivariate}
equalization/factorization effect of the model. This suggests the model(s) are actually operating under an information maximization goal.
However, this qualitative intuition has to be quantified.
To do so, we computed the average luminance and the RMSE contrast of $5\times10^6$ three-dimensional samples
extracted from the considered natural scenes, and we estimated their distribution in the luminance/contrast plane.
Then, these images were injected through the model: on the one hand the static Divisive Normalization response was computed,
and, on the other hand, the equivalent Wilson-Cowan system was stimulated with the linear frequency representation
of the brightness of each sample until it converged to a steady state (after 500 Euler integration steps).
Then, we computed the reduction in total correlation for the stimuli at different locations of the luminance/contrast plane.
We did this in two ways: (1) theoretically, through the proposed analytical expression,  Eq.~\ref{analitic}, and
(2) empirically, by using the RBIG method reviewed above, Eq.~\ref{eq_T_rbig}.
In each case, we considered up to $2.5\times10^4$ samples per location in the luminance/contrast plane
to do the estimations.

\begin{figure}[t]
 \hspace{-0cm} \includegraphics[width=1.0\textwidth]{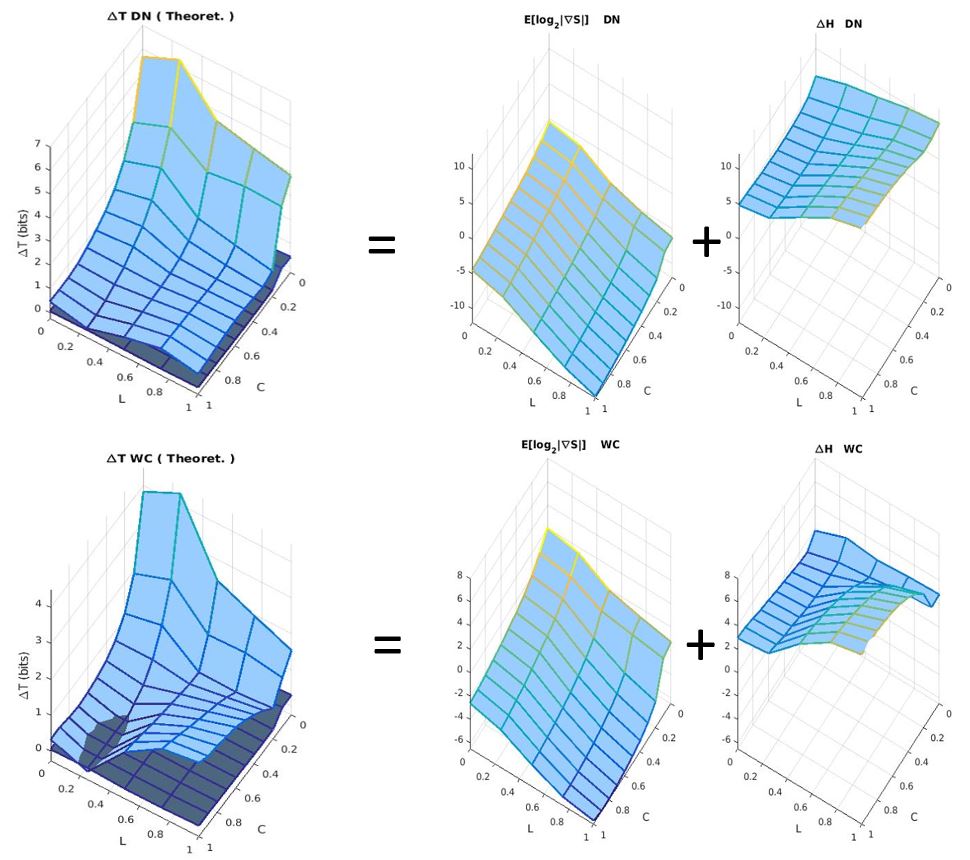}
 \caption{\small{\textbf{Theoretical efficiency of Wilson-Cowan and Divisive Normalization.} The theoretical reduction of total correlation for natural images of different luminance/contrast
 (surfaces at the left) is computed from a term that depends on the Jacobian of the transform (surfaces at the center), and a term that depends on the sum of marginal entropies of the input and output (surfaces at the right). The dark surfaces at the left are the (small) standard deviations of the estimates over 10 different realizations.}}
\label{theoretical}
\end{figure}

Fig.~\ref{theoretical} shows the elements of the theoretical computation: the final result comes from (i) a Jacobian-dependent term
(Eq.~\ref{analitic} in the Wilson-Cowan case, and Eqs. 42 and 43 in \cite{Martinez18} in the case of Divisive Normalization), and (ii)
and a term that depends on the marginal entropies.
As discussed above, the Jacobian-dependent term decreases with luminance and contrast, while the variation of the entropy-dependent
term is smaller across the image space. Therefore, the Jacobian determines the general behavior of the system.

\begin{figure}[t]
 \hspace{-0cm} \includegraphics[width=1.0\textwidth]{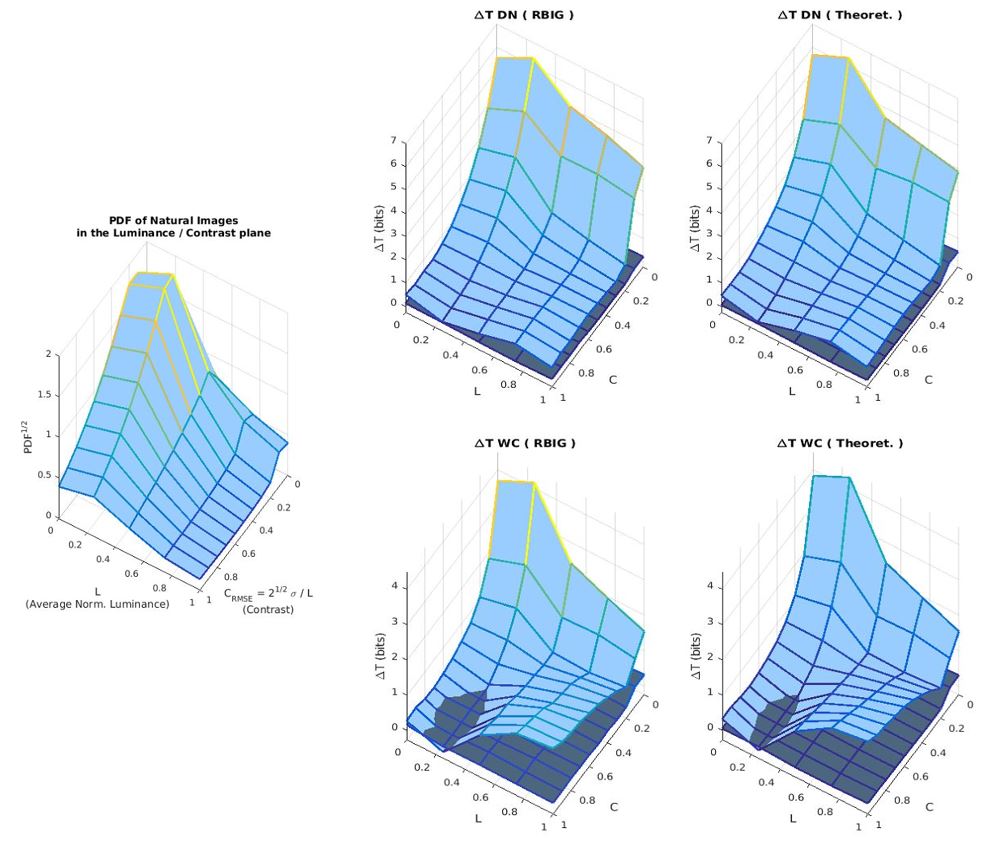}
 \caption{\small{\textbf{Efficiency is consistent with natural image statistics.} The general trend of the efficiency surfaces in the luminance/contrast plane for Divisive Normalization and Wilson-Cowan are the same. Moreover, both are consistent with natural image statistics since efficiency is bigger in the most populated region of the image space. The dark surfaces are the (small) standard deviations of the estimates over 10 different realizations.}}
\label{reduction_small_scale}
\end{figure}

Results of the theoretical and the empirical descriptions of the efficiency are compared in Fig.~\ref{reduction_small_scale} together with the estimated PDF in the luminance/contrast plane.
The general trend of the efficiency surfaces for Divisive Normalization and for the Wilson-Cowan model are the same (with both estimation approaches):
first, the redundancy is reduced all over the image space (the reduction of total correlation is positive almost everywhere), but more importantly,
the efficiency is clearly bigger for the low-luminance / low-contrast region, which also is the most populated region (see the PDF of the same
calibrated training set at the left).

These results confirm the accuracy of the empirical estimation of $T$ and the correctness of the theoretical expression for the Wilson-Cowan nonlinearity (or more precisely, they show the consistency between the theoretical and the empirical estimation).
Moreover, results show the equivalent behavior of Divisive Normalization and Wilson-Cowan in coding efficiency: both models focus on the same region of the image space. And, more interestingly, they show that psychophysically-inspired nonlinearities have their peak performance in the proper region of the image space even though they were not optimized in any way to that end.

\subsection{Empirical analysis of the full-scale model}
\label{empirical}

In this section we study the efficiency of the full-scale model using a purely empirical approach.
This is because the size of the Jacobian for each sample is huge and it is difficult to collect a big enough set for a reliable estimation of Eq.~\ref{analitic}.
For instance, in the current implementation of the model (using a 3-scale, 4-orientation, steerable transform \cite{Simoncelli92}),
discrete luminance patches of size $40\times40$ turn into response vectors $\vect{x} \in \mathbb{R}^{10025}$, so the Jacobian matrices are $10025\times10025$.
One could restrict the interest to a selected set of sensors (as done in some empirical analysis below). However, note that this restriction
does not imply a reduced-size Jacobian in the theoretical expression.
No rows/columns in the Jacobian can be neglected since the evolution of a subset of responses in the Wilson-Cowan integration depends on \emph{all} the responses
because, in principle, the matrix $\vect{W}$ is dense. And this dense nature makes perceptual sense because (stronger or weaker) there is interaction between all scales/orientations/positions~\cite{Foley94,Watson97}.

Here we analyze the behavior of the full-scale model in three different empirical ways:
(1)~by discussing the shape of the marginal PDFs for the responses of sensors at different layers of the network,
(2)~by computing the \emph{mutual information}, $I$, between the responses of multiple pairs of sensors at different layers (as in \cite{Malo10}),
and finally, and more interestingly,
(3)~by computing the \emph{total correlation}, $T$, among the responses at the different layers.
In the latter, different spatial sampling schemes are explored to capture how visual information depends on the field of view.

In all the experiments in this section we start from the responses of the considered model to the 5700 spectral
image patches described in the \emph{Materials} section. In each case, the responses at the different layers are
further subsampled according to the goal of the specific experiment and to get a representative set for the considered
estimation.

\subsubsection{Marginal equalization}

Descriptions of natural image statistics usually start from the marginal PDFs because
they show the basic complexity of the signal (e.g. naive coding would only consider zero-order entropy~\cite{Cover06}).
Marginal PDFs are relevant because once 2nd order correlation has been removed, marginal non-Gaussianity (or sparsity)
is an appropriate description of the total correlation in certain cases \cite{Cardoso03}.
It is also interesting to see how a perception system modifies the marginal PDFs (e.g. eventual equalization, increased sparsity, Gaussianization~\cite{Laughlin83,Ruderman94,Mumford99,Simoncelli98,Simoncelli01}) because
these changes may reveal an information maximization goal \cite{Laughlin83,Ruderman94} and the
general multivariate factorization goal \cite{Simoncelli01} may be achieved through more complicated
marginal PDFs~\cite{Malo10}.

The initial layers of the system we are considering (1st to 3rd layer) consist of sensors tuned to
specific spatial locations, while the 4th layer is formed by sensors with wavelet-like receptive fields.
In the marginal approach considered here we assume the signal is stationary across space and
orientation. Therefore we pool together samples from different spatial locations and (in the case of wavelets)
also corresponding to different orientations.
In all the results in this section we collected $1\times10^7$ randomly chosen samples to estimate each marginal PDF.

\begin{figure}[t]
 \hspace{-1.5cm} \includegraphics[width=1.2\textwidth]{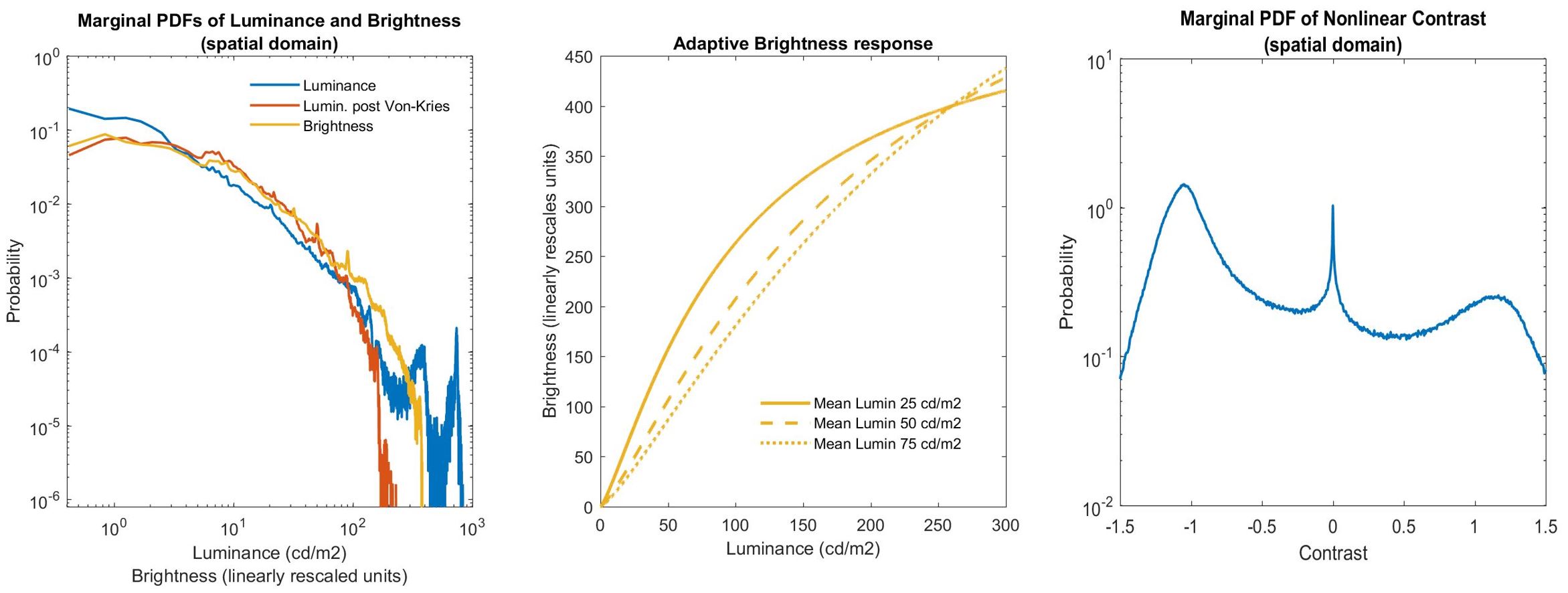}
 \caption{\small{\textbf{Marginal PDFs of responses of sensors tuned to spatial locations.} \emph{Left:} PDFs for the original luminance, Von-Kries luminance, and Brightness. \emph{Center:} Adaptive \emph{luminance to brightness} transform. \emph{Right:} PDF of nonlinear contrast. See  \cite{Martinez18} for details on specific Divisive Normalization expressions of the transforms.}}
\label{marginals_space}
\end{figure}

Figure~\ref{marginals_space} shows the effect of the layers of sensors tuned to spatial locations over the
original PDF of luminance at the retina.
We see that the linear luminance representation (blue line in Fig.~\ref{marginals_space}-left) is strongly biased and
has several peaks in the high-luminance tail. These different peaks correspond to high reflectance objects seen in
different illumination conditions.
Note how these peaks disappear after the Von-Kries normalization that puts every scene in the same relative-luminance
range. Then, the brightness transform that depends on the background luminance (Fig.~\ref{marginals_space}-center) tends
to expand the low-luminance range so that the resulting marginal PDF is relatively more flat.


All the above transforms make no use of spatial or contextual information (appart from the global scaling factors in Von-Kries
obtained from average illumination). The variation of the PDFs in these point-wise layers is consistent with equalization goals \cite{Laughlin83}.
This completely changes in the 2nd and 3rd layers that compute contrast from local normalizations of brightness and
apply linear center-surround receptive fields whose response is subsequently normalized by the local activity.

After contrast computation the mean is removed (see that the peak in Fig.~\ref{marginals_space}-right is in zero).
The distribution is not symmetric around zero because darker regions (now below zero) are more frequent in natural images.
Interestingly, the contrast masking through Divisive Normalization in the third layer generates a \emph{bimodal distribution}
above and below zero.
This effect in Divisive Normalization has been interpreted as predictive coding, where the numerator is predicted from the neighbors in the denominator leading to peaks above and below zero where this prediction is successful.
The predictive coding interpretation has been suggested many times \cite{Buccigrossi98,Epifanio00,Malo06a,Laparra16}, and peaks of this kind have been consistently found in \cite{Malo10,Martinez18} using different models and scenes.

\begin{figure}[t]
 \hspace{-1.5cm} \includegraphics[width=1.2\textwidth]{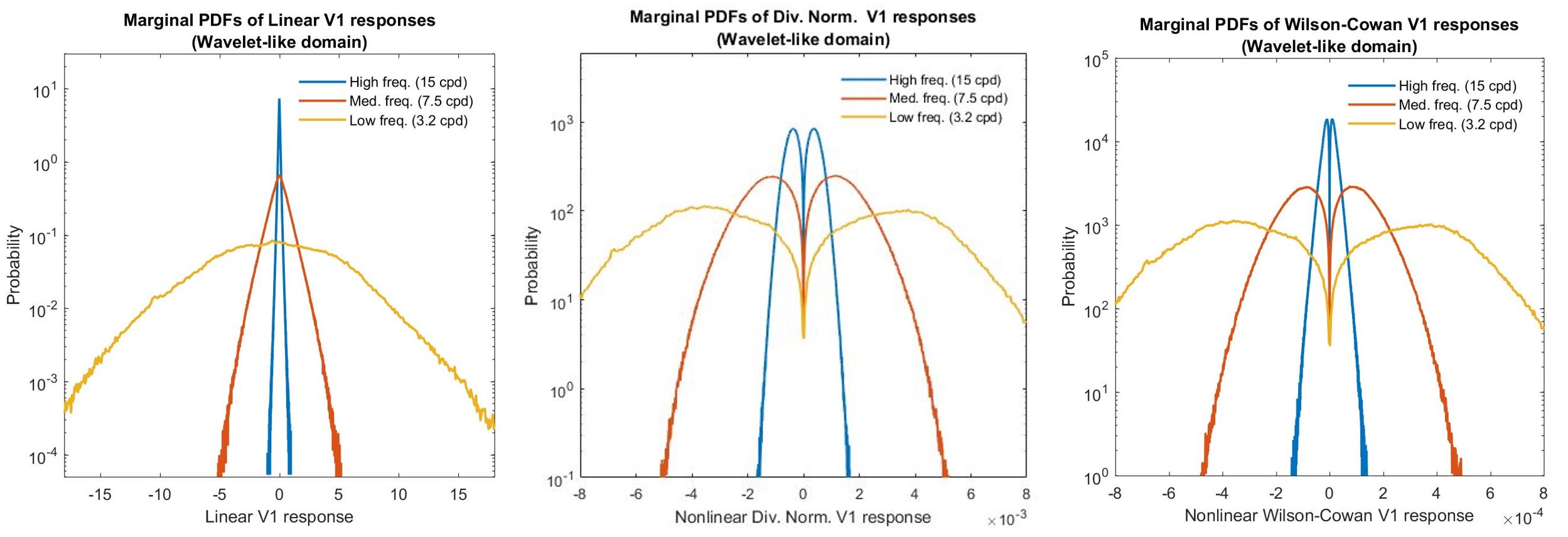}
 \caption{\small{\textbf{Marginal PDFs of responses of sensors tuned to local oriented features.}\emph{Left:}~PDFs~of the responses to distributions of nonlinear contrast of V1-like linear sensors. \emph{Center:}~PDFs~after Divisive Normalization. \emph{Right:}~PDFs~after Wilson-Cowan interaction.}}
\label{marginals_wavelet}
\end{figure}

Figure~\ref{marginals_wavelet} shows the marginal PDFs of the responses of sensors tuned to oriented features
of different scales. These sensors are the linear wavelet-like filters at the 4th layer that simulate V1
simple cells, and the corresponding mechanisms after the nonlinear interactions (either Divisive Normalization or Wilson-Cowan).
The linear wavelet-like sensors at the 4th layer display heavy tailed PDFs with decreasing variance for finer scales.
However the sparsity of these responses is lower than the sparsity of the same filterbank
applied on luminance images (results not shown).
This lower sparsity maybe because wavelet filters certainly lead to highly sparse (non-Gaussian) responses
when applied to luminance images, but the nonlinear-contrast images are substantially different
from luminance images (see for example the output of 3rd layer at Fig.~\ref{model}).

More interestingly, these (moderately) sparse response distributions turn into \emph{bimodal distributions} after the nonlinear interactions,
both for Divisive Normalization and Wilson-Cowan. Again (as in Fig.~\ref{marginals_space}-right) the masking interaction generates bimodal PDFs.
However, this is not a exclusive feature of Divisive Normalization: it happens in equivalent Wilson-Cowan systems as well.
Note that this marginal behavior (two modes around a depression in zero) can be understood in multivariate terms
using the visualization of the reduced-scale example in Fig.~\ref{manifolds}: the projection of the blobs on the different
axes leads to the peaks in the marginals.

\subsubsection{Mutual Information along the network}

Criticisms to linear Independent Component Analysis (ICA) pointed out that sparsity maximization
in the marginal PDFs does not guarantee complete statistical independence~\cite{Simoncelli01,Lyu09}.
This is obvious from the mathematical point of view~\cite{Cardoso03}, however,
in practice, a substantial amount of work was devoted to point out the existence of residual
statistical relations after ICA-like filters had been applied to images.
The problems for direct estimation of total correlation mentioned in the \emph{Methods} section,
impled a variety of surrogates to measure this remaining redundancy, as for instance the
analysis of conditional probabilities of neighbor responses, the so-called bow-ties~\cite{Buccigrossi98,Mumford99,Schwartz01,HyvarinenBubbles03,Malo06b,Malo10},
the analysis of the correlation between the energies of pairs of responses as a way to identify subspaces with residual relations~\cite{HyvarinenTICA,HyvarinenISA,Hyvarinen09,Epifanio00,Epifanio03},
or the measure of mutual information between pairs of responses~\cite{Moulin01,Malo06a,Malo10}.

Here we quantify the relation between pairs of responses using mutual information,~$I$, as a function of the separation between
the sensors in the corresponding feature space (spatial departure for the first three layers and departure in space, scale and orientation for the wavelet-like layer).
We applied the straightforward definition of mutual information because each computation reduces
to the estimation of a joint (bivariate) PDF and two marginal PDFs.
We did that by gathering neighbors from sliding windows of size smaller than the available visual field.
In that way we got $8\times10^4$ samples for each pair of sensors. We computed 10 estimations of $I$ using a randomly chosen subset with 80\% of
those samples in each estimation. We chose the bin size according to the Silverman rule of thumb \cite{Silverman86}.
The results in the plots are the average of those estimations.

\begin{figure}[t]
 \hspace{-1.5cm} \includegraphics[width=1.2\textwidth]{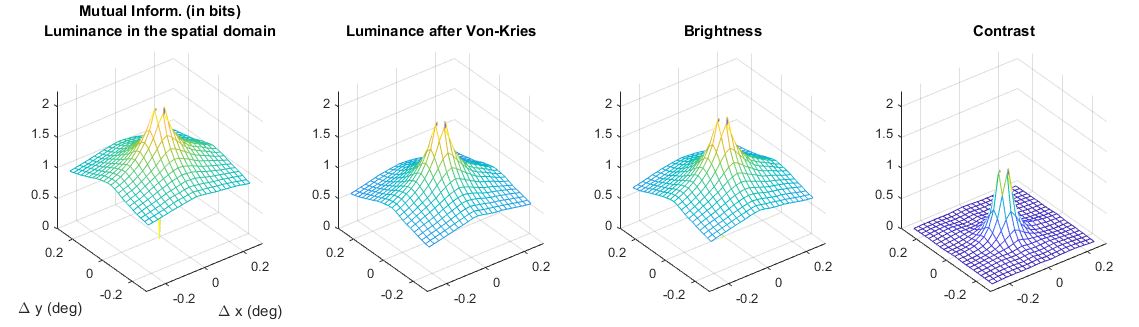}
 \caption{\small{\textbf{Mutual information between a spatial sensor and its neighbors.} Different layers along the network are considered from left to right. The value of auto-mutual information has been set to zero for better visualization. We did that because this value is arbitrarily large (every response contains arbitrarily large amount of information about itself), and it should not be considered in describing the interaction with the neighbors.}}
\label{mutuals_space}
\end{figure}

Figure~\ref{mutuals_space} shows the evolution of the interactions as the signal goes through
the three initial layers.
In every case, the relation is higher with closer neighbors and decays with distance.
However, in the explored spatial range (which is limited, about 0.5 degrees)
these pair-wise relations do not drop to zero in layers before contrast computation.
The $I$ would eventually arrive to zero for sufficiently large separation, but
we didn't have access to big enough distances.
Von-Kries adaptation and brightness computation reduce the amount of paired relations about 0.5 bits, but the
substantial change again comes with the introduction of spatial processing in the third layer: the relations
with the closer neighbors are smaller, they drop faster, and actually arrive to zero for small departures.
The division by the local activity removes the relation with the spatial neighborhood: it seems that
the predictive coding interpretation of normalization mentioned above \cite{Buccigrossi98,Epifanio00,Malo06a,Laparra16}
also applies here.

\begin{figure}[t]
 \hspace{-1.5cm} \includegraphics[width=1.2\textwidth]{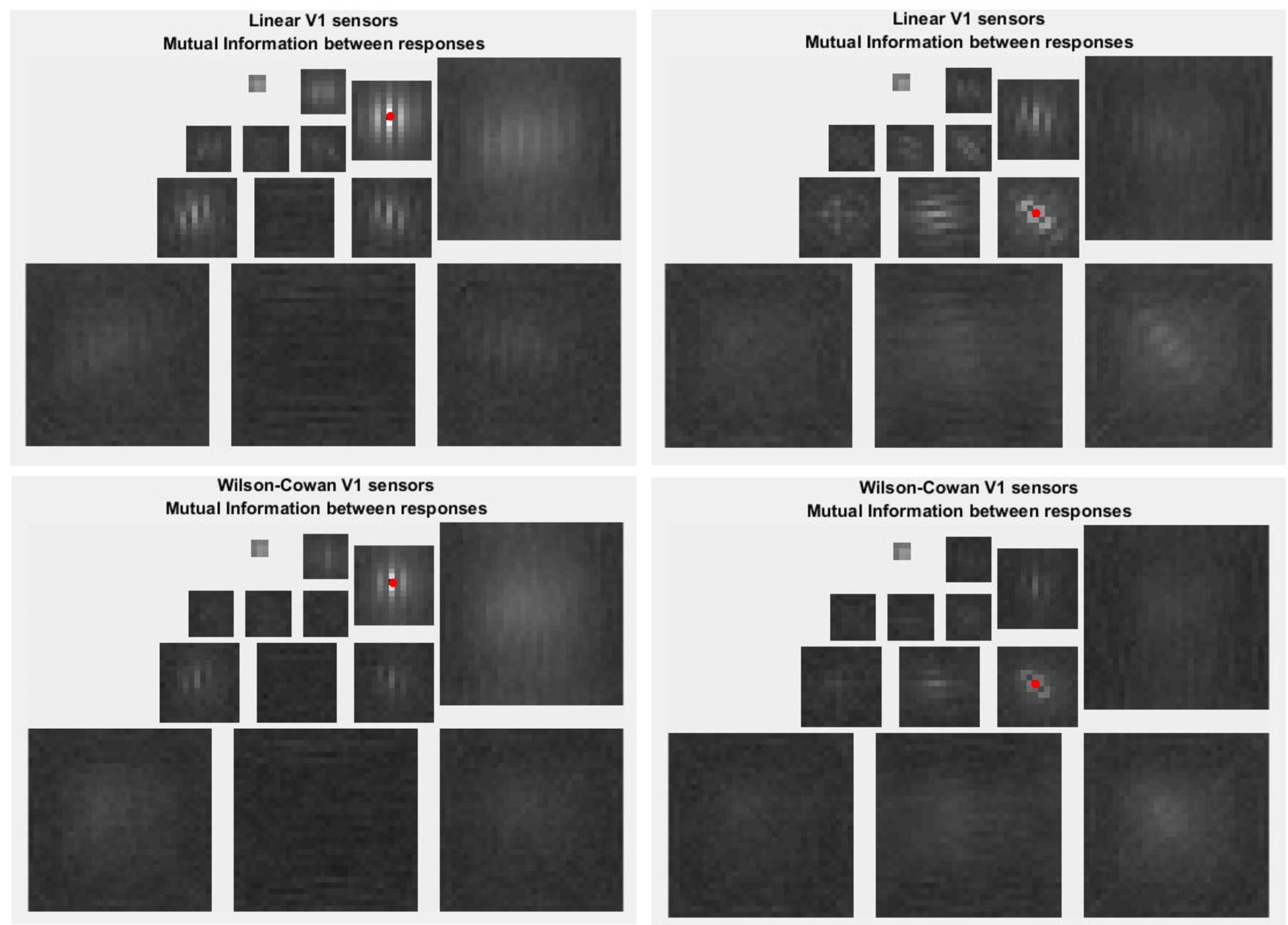}
 \caption{\small{\textbf{Mutual information between two wavelet-like sensors and all their neighbors.} \emph{Left:} interactions of a sensor tuned to vertical patterns of middle frequency (highlighted in red). \emph{Right:} interactions of a sensor tuned to diagonal patterns of middle frequency (highlighted in red). \emph{Top:} interactions between the linear sensors. \emph{Top:} interactions between the sensors after the Wilson-Cowan recurrence. Lighter gray indicate higher mutual information. All figures are scaled in the same way. Comparing \emph{top} and \emph{bottom} one can see that the Wilson-Cowan recurrence reduces the statistical relations a little bit (darker diagrams at the bottom).}}
\label{mutuals_wavelet1}
\end{figure}

\begin{figure}[t]
 \hspace{2.5cm} \includegraphics[width=0.7\textwidth]{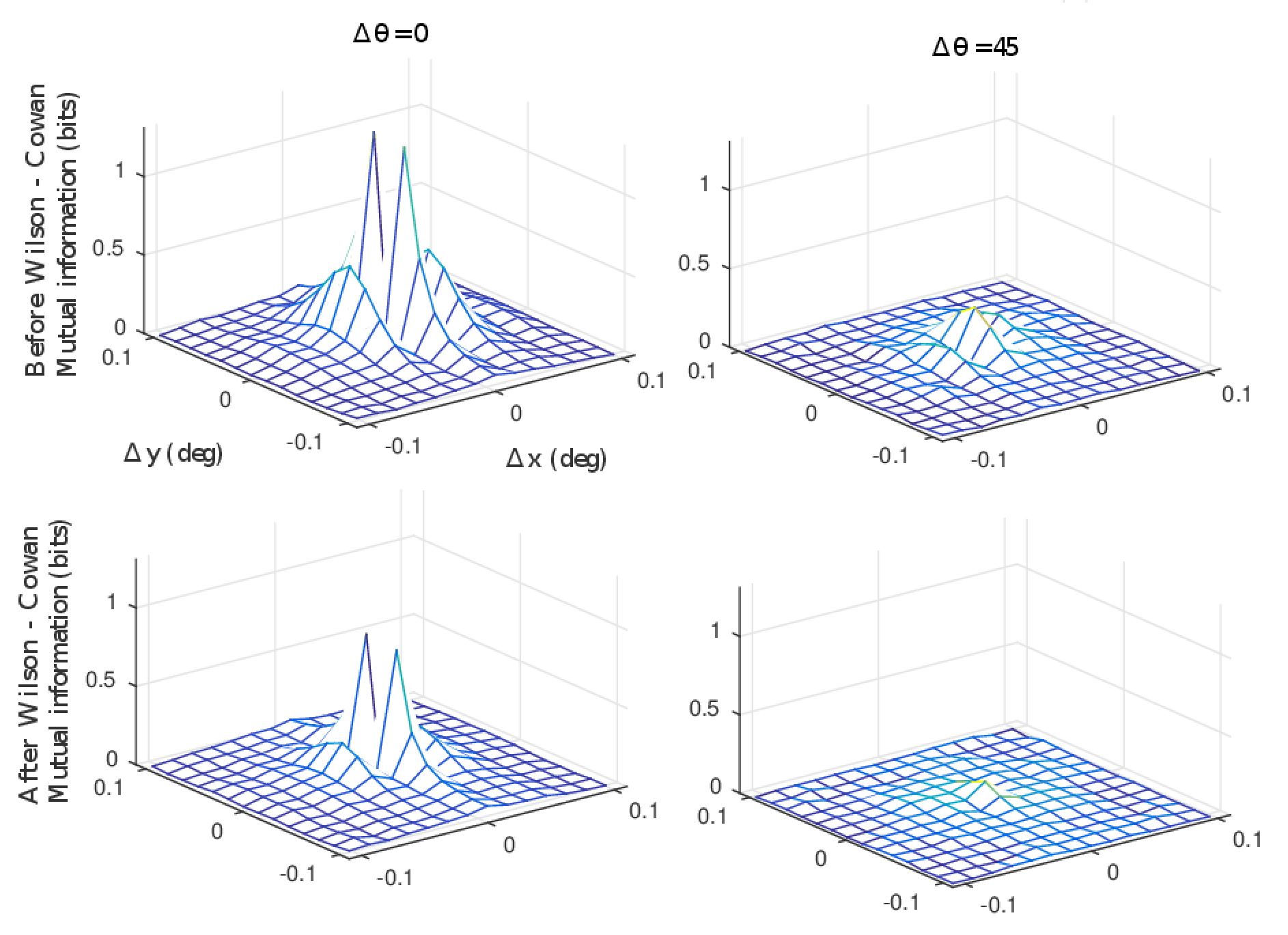}
 \caption{\small{\textbf{Mutual information between a wavelet-like sensor and its neighbors in space and orientation.} The considered sensor is at the center of the domains represented at the left column. \emph{Left:} Decay of the interaction in space (for the same orientation -or subband-). \emph{Right:} Decay of the relation in space for a different orientation. \emph{Top:} result before the Wilson-Cowan recurrence. \emph{Bottom:} result after the Wilson-Cowan recurrence.}}
\label{mutuals_wavelet2}
\end{figure}

Figure~\ref{mutuals_wavelet1} shows the mutual information values for the responses of two sensors (those highlighted in red)
with all their neighbors in the response vector at the V1 layer before and after the application of the Wilson-Cowan interaction.
Finally, Fig.~\ref{mutuals_wavelet2} shows a subset of the results shown in Fig.~\ref{mutuals_wavelet1} represented as a surface
as in Fig.~\ref{mutuals_space}.
As previously reported for linear wavelet domains \cite{Moulin01,Laparra10b,Malo10}, here mutual information also decays with distance in space
scale and orientation. After the nonlinear interactions (we only show the Wilson-Cowan result) the relations between the coefficients seem to be reduced.

However, the reduction obtained here is substantially smaller than the one obtained with previous Divisive Normalization models where $I$ was computed in the same way~\cite{Malo10}.
It is not a specific matter of Wilson-Cowan, but it is related to the multi-layer architecture: results with the current architecture with Divisive Normalization (not shown) are similar to the Wilson-Cowan results in Figs.~\ref{mutuals_wavelet1} and~\ref{mutuals_wavelet2}, i.e. only moderate gain in~$I$. It is not the database either: we made preliminary experiments with the current architecture and Divisive Normalization using the Van Hateren stimuli \cite{VanHateren98} (the data used in~\cite{Malo10}), and the results (not shown) are similar: also a very moderate gain.

The difference in the architecture explored here with regard to the one in~\cite{Malo10} is that the current one is \emph{deeper}: the current wavelet-like layer comes after previous layers that already are doing spatially significant Divisive Normalization (in our case the 3rd layer).
The presence of previous layers which already reduce the redundancy, may limit the ability of the considered nonlinearities at this later stage. Nevertheless, note that the set of mutual information measures is only a surrogate
of the conceptually appropriate measure, the total correlation, which is addressed in the next section.

\subsubsection{Total Correlation along the network}

In this section we measure the shared information (total correlation) in the responses of sensors covering progressively bigger portions of the visual field, and we see how this shared information modifies along the neural pathway. We take two different spatial sampling strategies: (1) \emph{no-subsampling}, i.e. considering \emph{all} the sensors tuned to the spatial region, so bigger regions imply a bigger number of sensors, and (2) \emph{subsampling}, i.e. taking a fixed number of sensors progressively more separated, so the bigger the visual field, the poorer the sampling of the visual patterns.

\begin{figure}[b]
 \hspace{-1.5cm} \includegraphics[width=1.2\textwidth]{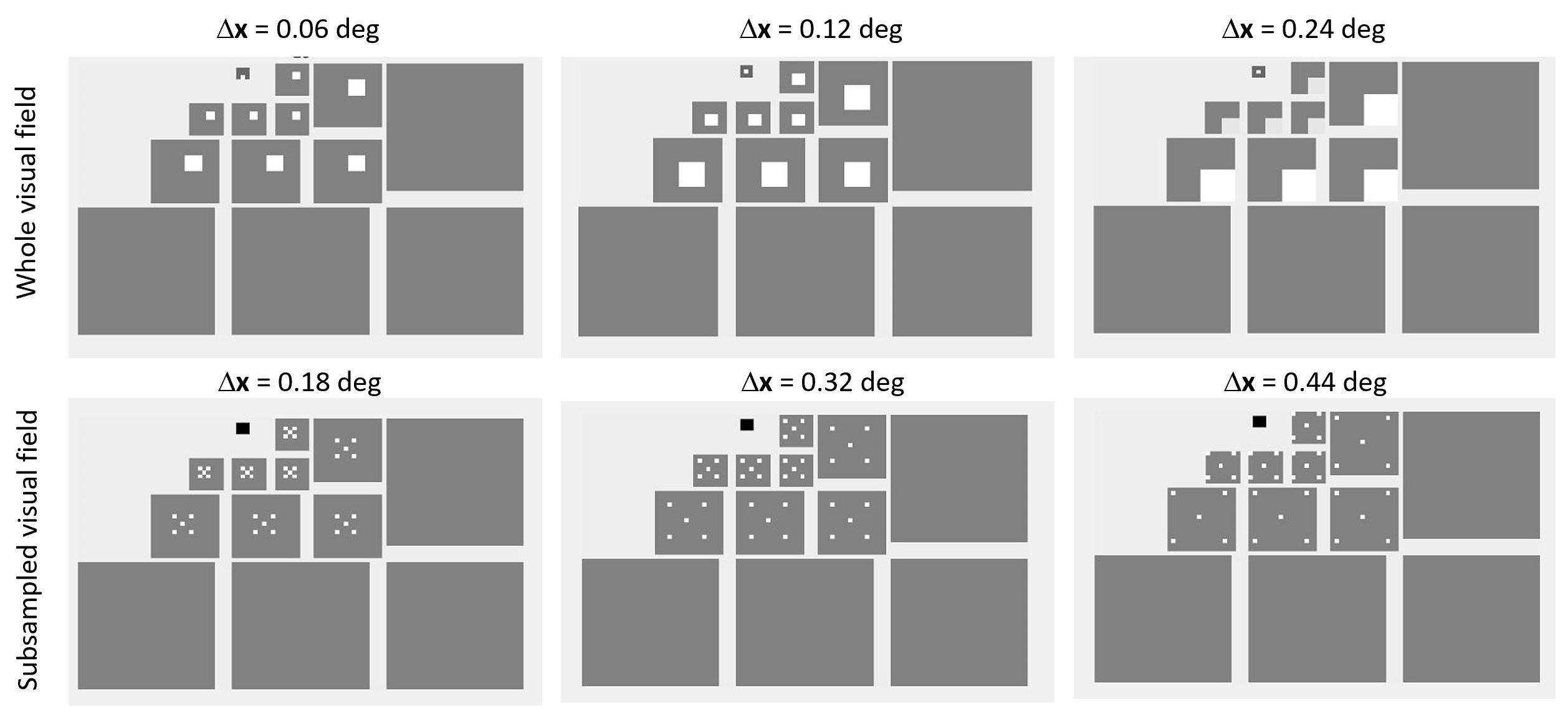}\\\vspace{-0.5cm}
 \caption{\small{\textbf{Wavelet sensors in total correlation experiments.}
 Visual angle and spatial sampling strategy. From left to right: increasing visual angle covered by the sensors. Top and bottom: no sub-sampling, versus sparse sampling.}}
\label{sensors_wavelet}
\end{figure}

An illustration of the dimensions of the response vector considered at the wavelet-like layer in both experiements is shown in Fig.~\ref{sensors_wavelet}.
The corresponding selection for the layers where sensors have spatial meaning is straightforward.

Taking such subsets of responses (or sensors) keeps the dimension small and makes the empirical estimation easier.
This specific sampling schemes were proposed to allow gathering response vectors by sliding windows like those in Fig.~\ref{sensors_wavelet}.
Collecting big datasets is particularly important since the dimensions of the considered vectors range from 5 (in the \emph{subsampling experiment} in the layers with spatial meaning) up to 321 (in the \emph{no-subsampling experiment} with bigger field of view in the wavelet-like layers). In every case, the total correlation is reported \emph{by coefficient}.
All the total correlation measures in this section were computed using $2.5\times10^4$ samples and
RBIG with 600 layers to ensure the convergence to the Gaussian. In each case, 10 realizations of the estimates were computed using $80\%$ of the samples.

\begin{figure}[b]
 \centering
 \vspace{0cm}
 \begin{tabular}{cc}
 \hspace{-1.5cm}\includegraphics[height=0.55\linewidth]{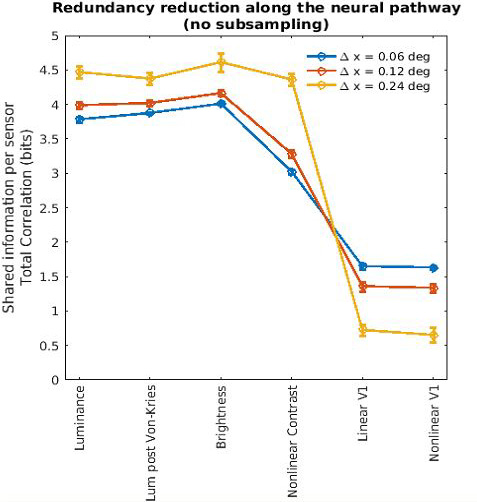} & \includegraphics[height=0.55\linewidth]{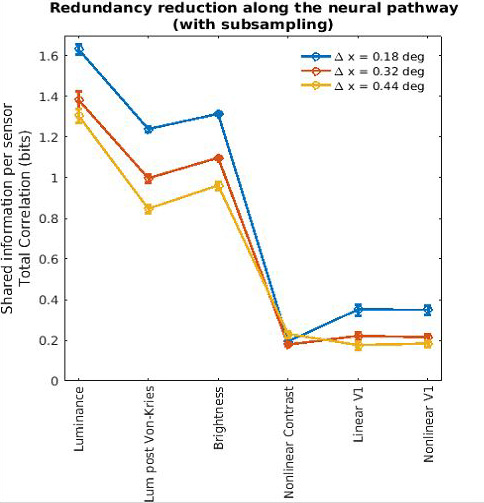}\\[-0.2cm]
 \end{tabular}
 \caption{\small{\textbf{Reduction of Total Correlation along the layers of the network.} Visual signals of different spatial size are considered. \emph{Left}: no spatial subsampling (actual images).  \emph{Right}: with spatial subsampling (samples at the corners and the center of the visual field).}}
 \label{reduction_of_T}
 \vspace{-0.3cm}
\end{figure}

The sampling strategies were selected because they lead to a clear pattern in the total correlation in the input representation.
This pattern is a safety check in this full-scale situation where the theoretical computation is not feasible due to the size of the Jacobians.
Note that given the smoothness of objects in natural scenes, the consideration of bigger visual fields in the \emph{no-subsampling experiment} should include
more coherent structures (e.g. bigger portions of objects in the images). As a result, one would expect increased redundancy given the bigger coherence of the visual structures.
On the other hand, if a fixed number of sensors is progressively separated covering bigger visual fields, as in the \emph{subsampling experiment}, the statistical relation between the sensors in the extremes of the visual field should decrease with separation.

This is exactly the pattern that is found in the results, see Fig.~\ref{reduction_of_T}: (1) in luminance images in
the \emph{no-subsampling experiment}, total correlation increases with the size of the visual field, and (2) in luminance images in
the \emph{subsampling experiment}, total correlation decreases with the size of the visual field.
Note also that the stimuli that can be properly called \emph{images} are those of the \emph{no-subsampling experiment}.
In the \emph{subsampling experiment} we just take samples at the corners and at the center of the visual field.

Regarding the efficiency of the signal representations, Fig.~\ref{reduction_of_T}-left shows how the total correlation is reduced for \emph{images}
(note that the signals in the \emph{subsampling experiment} are not strictly \emph{images}).
Results show that the initial layers (Von-Kries chromatic adaptation, and Weber-like
adaptive brightness) do not contribute to redundancy reduction in images.
This is because these operations introduce no fundamental spatial information, so
the redundancy between neighbor spatial locations is basically preserved.
Redundancy starts to be removed in the third layer (\emph{Nonlinear Contrast})
where the Contrast Sensitivity Function is applied to the local contrast and then the result is nonlinearly
transformed using Divisive Normalization.
In this layer, bigger reductions are obtained in smaller visual fields due to
the (relatively small) size of the masking kernel of Divisive Normalization in the spatial domain (about 0.02 degrees).
On the contrary, the bigger the visual field (which involves more complicated visual structures), the more effective is the linear filterbank of wavelet-like receptive fields (\emph{Linear V1} layer).
Finally, the nonlinear interaction at the cortical level (after all the previous stages)
does not imply substantial reductions of redundancy in terms
of total correlation. The values at the \emph{Nonlinear V1} layer in Fig.~\ref{reduction_of_T} display the specific result
for the Wilson-Cowan interaction, but Table~\ref{table_TC} shows that the removed redundancy for the equivalent
Divisive Normalization model is virtually the same: in the biggest visual field explored for images the inner representation
removes about 85\% of the total correlation in the input representation.

Fig.~\ref{reduction_of_T}-right shows the equivalent results for visual fields in which spatial samples were taken from
the corners and the center of the considered regions. In this configuration the inner representation also removes about 85\% of the
total correlation at the input. However, in this case, most of the redundancy reduction comes from the nonlinear
contrast computation.

Similarly to what was found in the mutual information analysis, in this cascade (for images) the biggest reduction in redundancy
comes from the wavelet-like transform and the contribution of the nonlinear interaction between simple cells is
relatively small.

\begin{table}[t!]
    \caption{\small\textbf{Removed Redundancy (in $\%$).} Percentage of Total Correlation removed at the inner representation with regard to the values at the input representation. Visual fields of different size (in degrees) with no subsampling (left) and with subsampling (right).}\label{table_TC}
    \vspace{-0.3cm}
    \hspace{-1cm}
\begin{tabular}{|l|ccc|ccc|}
  \hline 
                   &   & NO-SUBSAMPL. &  &  & SUBSAMPL.  &  \\
                 & $\Delta x = 0.24$ & $\Delta x = 0.12$ & $\Delta x = 0.06$ & $\Delta x = 0.44$ & $\Delta x = 0.32$  & $\Delta x = 0.18$ \\\hline
  Divisive Norm. & $\mathbf{84 \pm 3}$ & $67\pm3$ & $54\pm2$ & $87\pm2$ & $84\pm4$ & $79\pm3$ \\
  Wilson-Cowan   & $\mathbf{85 \pm 4}$ & $67\pm3$ & $54\pm1$ & $86\pm3$ & $84\pm3$ & $79\pm3$ \\
  \hline
\end{tabular}

\end{table}


\section{Discussion}

\paragraph{Alternative confirmation of Barlow's hypothesis for Wilson-Cowan systems.}
In this work we performed the first analysis of the communication efficiency of Wilson-Cowan networks in accurate information theoretic terms
using the appropriate (multivariate) description of redundancy: the total correlation.
Pointing out the efficiency of a psychophysically tuned system is an alternative confirmation of Barlow's \emph{Efficient Coding Hypothesis}.

This hypothesis states that natural perception systems evolved to efficiently encode natural signals~\cite{Barlow59,Barlow61,Barlow01}.
The conventional way to check this hypothesis is training artificial systems with natural scenes assuming some communication efficiency goal
(as for instance redundancy reduction), and see the emergence of natural properties from the statistical learning.
This logic (\emph{from-statistics-to-perception}) has been applied to explain the linear image representation in V1 \cite{Olshausen96,Hyvarinen09,Gutmann14},
the nonlinear interactions~\cite{Schwartz01,Malo06b,MacLeod03b,Laparra12}, and even the origin of visual illusions~\cite{Barlow90,Laparra15}.

However, note that here we reasoned in the opposite direction: we took a psychophysically plausible network and analyzed its behavior
in information-theoretic terms after no statistical learning whatsoever.
We found that the considered systems have bigger efficiency (bigger reductions in $T$) in the most populated regions of the
image space, see Fig.~\ref{reduction_small_scale}, and they remove about 85\% of the total correlation of natural scenes in the input representation, see Fig.~\ref{reduction_of_T} and Table~\ref{table_TC}.
The approach used here (\emph{from-perception-to-statistics}) is an alternative way to confirm the Efficient Coding Hypothesis, as suggested in~\cite{Malo10}.

\paragraph{Efficiency of Wilson-Cowan is similar to Divisive Normalization.}
Several evidences suggest that the Wilson-Cowan interaction can be as effective in information theoretic terms as the Divisive Normalization model.
First, consider that the interaction kernel $\vect{W}$ derived here from the relation proposed in \cite{Malo18,Malo19}, roughly decays with space, orientation, and scale. The similarity between the interaction kernel and the mutual information in the linear wavelet-like representation (compare Figs.~\ref{parameters_WC}-right and~\ref{mutuals_wavelet1}-left) has been related to the effectiveness of this kind of interactions to capture
the structure of the signal~\cite{Laparra10b,Malo10}.
Second, recurrent subtraction of the activity of the neighbors weighted in this way in the integration of the Wilson-Cowan equation lead to similar \emph{bimodal} marginal PDFs, and to similar clusters in the joint PDF as the Divisive normalization (see Figs.~\ref{marginals_wavelet} and~\ref{manifolds}).
Third, the Jacobian of the Wilson-Cowan nonlinearity induces similar trends in the reduction of total correlation as the Divisive Normalization
(see Figs.~\ref{theoretical} and~\ref{reduction_small_scale}): both models have better efficiency in the low-luminance / low-contrast region.
And finally, note that the redundancy reduction in the full-scale model is roughly the same using the Wilson-Cowan interaction and the Divisive Normalization (see Table \ref{table_TC}).

The above statistical similarities between the Wilson-Cowan and the Divisive Normalization models
are consistent with the ability of Wilson-Cowan interaction to improve the explanation of pattern visibility, which is similar to Divisive Normalization.
Note that the results in Fig.~\ref{correlation} with Wilson-Cowan are consistent with better explanations of subjective image quality reported for Divisive Normalization \cite{Laparra10a,Laparra17,Martinez18}.
These information theoretic and perceptual similarities suggest the Divisive Normalization actually is the steady state of the Wilson-Cowan
dynamics, as assumed in~\cite{Malo19}.

\paragraph{Accuracy of total correlation estimates.} A side technical benefit of the theory presented here
(the expression of the reduction of total correlation for Wilson-Cowan systems, Eqs.~\ref{eq_T_studeny} and~\ref{analitic})
is that this kind of theoretical results can be used to check the accuracy of empirical estimates of total correlation in neuroscience applications.
This is also true for the equivalent result for Divisive Normalization presented in~\cite{Martinez18}.
Here we directly compare these theoretical results with empirical estimates (in Fig.~\ref{reduction_small_scale}),
which was not done in~\cite{Martinez18}.

These ground truth results are relevant because information measures using Gaussianization \cite{Laparra11,ICLR19} had been used to identify interesting patterns in other spatio-spectral and spatio-temporal data~\cite{Johnson2018,Laparra15igarss}, but in
those cases there was no theoretical bound to compare with.

The success of the empirical measure of total correlation \cite{Laparra11,ICLR19} shown in this work suggests that  it can be used as
a goal function or as an analysis tool to optimize artificial systems, or to check natural systems according to the
information bottleneck principle \cite{Tishby18}.

%

\paragraph{Differences with shallower networks.}
Despite here we quantify redundancy through a more appropriate concept, the analysis presented in this work is similar
to the one done in~\cite{Malo10} for a shallower network based on another phychophysically tuned Divisive Normalization.
The model consisted of a single linear+nonlinear layer:
a wavelet transform plus a linear weighting to simulate the contrast sensitivity followed by a Divisive Normalization
of the energies of the weighted wavelet.

However, as noted in section \ref{empirical}, the results obtained here differ from those
reported in~\cite{Malo10}: the reduction of the mutual information with regard to the linear wavelet representation
obtained here is substantially smaller. This is consistent with the small reduction in total correlation with regard to the linear wavelet representation shown in Fig.~\ref{reduction_of_T}, which was not measured in~\cite{Malo10} because the statistical tool was not available.
This could mean that the efficiency of this interaction (either implemented through Wilson-Cowan or Divisive Normalization)
is not as big as suggested in~\cite{Malo10}.

On the contrary, we argue that the small gain with regard to the linear wavelet is not a fundamental limitation of the Divisive Normalization
or the Wilson-Cowan models, but a simple consequence of the fact that we are looking at a \emph{deeper} layer:
the presence of previous layers which already reduce the redundancy limits the ability of the considered nonlinearities a later stage.
In fact, the reduced sparsity displayed by our linear wavelet representation in Fig.~\ref{marginals_wavelet}-left has the same origin:
the modified effect of wavelets on contrast images (instead of the conventional luminance images).

%

\paragraph{Reproducible results.}
Code to reproduce all the empirical results on the behavior of the full-scale model as well as the experiment of the reduced-scale model that includes the theoretical result are available at \texttt{https://github.com/alviur/information\_wc.git}

\paragraph{Final remarks.}
In summary, the considered psychophysically tuned 4-layer network removes 85\% of the total correlation from
achromatic (luminance) images subtending 0.24~degrees.
For this field of view the Wilson-Cowan model has the same communication efficiency as the equivalent Divisive Normalization model.
In a reduced-scale scenario, the theoretical and the empirical results on redundancy reduction are consistent, and
both show that the psychophysically inspired Wilson-Cowan and Divisive Normalization networks are more efficient
in the regions of the image space where natural scenes are more frequent.
The above results represent a confirmation of Barlow's Efficient Coding Hypothesis for Wilson-Cowan models
in the \emph{perception-to-statistics} direction (alternative to the conventional \emph{statistics-to-perception} approach).

The similarities in the statistical effect of the Wilson-Cowan and the Divisive Normalization suggest
that recurrent neural field models could be an alternative in image coding
applications where Divisive Normalization has beaten both JPEG and JPEG2000~\cite{Malo06a,ICLR17}.
From a more fundamental point of view, future work should use the proposed analysis before the stationary state is reached. This will show how the information efficiency may evolve while adaptation takes place.

\appendix
\section{Appendix: Psychophysically inspired reduced-scale model}
\label{reduced-scale}

\paragraph{Overview.} The reduced-scale model consist of two \emph{linear+nonlinear} layers: (1) a linear \emph{radiance-to-luminance} transform using a standard Spectral Sensitivity Function, $V_\lambda$,
in the spectral integration~\cite{Stiles82}, followed by a simple exponential for the \emph{luminance-to-brightness} nonliniearity applied pixel-wise in the spatial domain, that simulates the Weber-Fechner response to luminance~\cite{Fairchild13},
and (2) a \emph{linear+nonlinear} layer in which the linear transform is a discrete cosine transform (a orthonormal rotation) followed by a
low-pass weighting function that simulate frequency-tuned sensors and the
Contrast Sensitivity Function (CSF) \cite{Campbell68}. Then, the outputs of the frequency sensors undergo a nonlinear interaction
that may be a Divisive Normalization~\cite{Carandini94,Martinez18,Martinez19}, or its equivalent
Wilson-Cowan network, with parameters computed according to Eq.~\ref{relation_W_H}~\cite{Malo19}.

\begin{equation}
  \xymatrixcolsep{2pc}
  \xymatrix{ \vect{x}^0 \ar@/^1pc/[r]^{\scalebox{1.00}{$\mathcal{L}^{(1)}$}} & \vect{r}^1  \ar@/^1pc/[r]^{\scalebox{1.00}{$\mathcal{N}^{(1)}$}} & \vect{x}^1 \ar@/^1pc/[r]^{\scalebox{1.00}{$\mathcal{L}^{(2)}$}}  & \vect{r}^2 \ar@/^1pc/[r]^{\scalebox{1.00}{$\mathcal{N}^{(2)}$}} & \vect{x}^2
  }
  \label{modular}
\end{equation}

\paragraph{Transform.} The actual inputs of our code are the responses of the linear photoreceptors: 3-pixel image vectors with normalized luminance values, i.e. $\vect{r}^1 \in \mathbb{R}^3$.
The normalized luminance was computing dividing the absolute luminance in $cd/m^2$ by the value corresponding to the 95\% percentile
of the luminance, in our case 260 $cd/m^2$.

\begin{itemize}

\item The \emph{luminance-to-brightness} transform, $\mathcal{N}^{(1)}$, is just:
\begin{equation}
      \vect{x}^1 = (\vect{r}^1)^\gamma \,\,\,\,\,\,\,\,\,\, \textrm{where} \,\,\,\, \gamma = 0.6
\end{equation}

\item The linear transform of frequency-tuned sensors with CSF gain, $\mathcal{L}^{(2)}$, is:
\begin{equation}
      \vect{r}^2 = G_{\textrm{CSF}} \cdot F \cdot \vect{x}^1 \,\,\,\,\,\,\,\,\,\, \textrm{where}
\end{equation}
\vspace{-0.5cm}
\begin{eqnarray*}
  F &=& \left(
    \begin{array}{ccc}
      \sqrt{\frac{1}{3}}  & \sqrt{\frac{1}{3}} & \sqrt{\frac{1}{3}} \\[0.3cm]
      \sqrt{\frac{1}{2}}  & 0 & - \sqrt{\frac{1}{2}} \\[0.3cm]
      -\sqrt{\frac{1}{6}} & \sqrt{\frac{2}{3}} & -\sqrt{\frac{1}{6}} \\
    \end{array}
  \right)\\[0.3cm]
  G_{\textrm{CSF}} &=& \left(
    \begin{array}{ccc}
     \,\, 1 \,\,\,\, & \,\,\,\, 0 \,\,\,\,& \,\,\,\, 0 \,\, \\
     \,\, 0 \,\,\,\, & \,\,\,\, 0.5 \,\,\,\,& \,\,\,\, 0 \,\, \\
     \,\, 0 \,\,\,\, & \,\,\,\, 0 \,\,\,\,& \,\,\,\, 0.3 \,\,
    \end{array}
  \right)
\end{eqnarray*}

\item The \emph{Divisive Normalization} of the frequency-tuned sensors, $\mathcal{N}^{(2)}_{\textrm{DN}}$, is:
\begin{equation}
    \vect{x}^2 = sign(\vect{x}^2) \odot \mathbb{D}_{\vect{k}} \cdot \mathbb{D}^{-1}_{\left( \vect{b} + \vect{H} \cdot |\vect{r}^2|^\gamma \right)} \cdot |\vect{r}^2|^\gamma  \,\,\,\,\,\, \textrm{where} \,\,\,\,\,\, \gamma = 0.7, \,\,\,\,\,\,  \textrm{and,}
    \label{DN_B2}
\end{equation}
\vspace{-0.5cm}
\begin{eqnarray*}
  \mathbb{D}_{\vect{k}} &=& \left(
    \begin{array}{ccc}
     \, 0.18 \,\,\, & \,\,\, 0 \,\,\,& \,\,\, 0 \, \\
     \, 0 \,\,\, & \,\,\, 0.03 \,\,\,& \,\,\, 0 \, \\
     \, 0 \,\,\, & \,\,\, 0 \,\,\,& \,\,\, 0.01 \,
    \end{array}
  \right)\\
  H &=& \mathbb{D}_l \cdot W \cdot \mathbb{D}_r = \left(
    \begin{array}{ccc}
     \, 0.06 \,\,\, & \,\,\, 0    \,\,\,& \,\,\, 0 \, \\   
     \, 0    \,\,\, & \,\,\, 0.35 \,\,\,& \,\,\, 0 \, \\
     \, 0    \,\,\, & \,\,\, 0    \,\,\,& \,\,\, 0.27 \,
    \end{array}
  \right)
  \cdot
  \left(
    \begin{array}{ccc}
     \, 0.93 \,\,\, & \,\,\, 0.06 \,\,\,& \,\,\, 0.01 \, \\
     \, 0.04 \,\,\, & \,\,\, 0.93 \,\,\,& \,\,\, 0.05 \, \\
     \, 0    \,\,\, & \,\,\, 0.02 \,\,\,& \,\,\, 0.98 \,
    \end{array}
  \right)
  \cdot
  \left(
    \begin{array}{ccc}
     \, 0.95 \,\,\, & \,\,\, 0    \,\,\,& \,\,\, 0    \, \\
     \, 0    \,\,\, & \,\,\, 0.27 \,\,\,& \,\,\, 0    \, \\
     \, 0    \,\,\, & \,\,\, 0    \,\,\,& \,\,\, 0.13 \,
    \end{array}
  \right)
\end{eqnarray*}
and the vector of semisaturations, $\vect{b}$, is:
\begin{equation*}
\vect{b} = \left(
  \begin{array}{c}
    0.08 \\ 
    0.03 \\ 
    0.01 \\ 
  \end{array}
\right)
\end{equation*}

\item The equivalent Wilson-Cowan interaction, $\mathcal{N}^{(2)}_{\textrm{WC}}$, is defined by the differential equation \ref{EqWC}, where the auto-attenuation, $\vect{\alpha}$, and the interaction matrix, $\vect{W}$, are:
\begin{eqnarray}
\vect{\alpha} &=& \left(
  \begin{array}{c}
    0.41 \\ 
    1.10 \\
    1.30 \\
  \end{array}
\right)\\
  W &=& \left(
    \begin{array}{ccc}
     \, 0.93 \,\,\, & \,\,\, 0.06 \,\,\,& \,\,\, 0.01 \, \\
     \, 0.04 \,\,\, & \,\,\, 0.93 \,\,\,& \,\,\, 0.05 \, \\
     \, 0    \,\,\, & \,\,\, 0.02 \,\,\,& \,\,\, 0.98 \,
    \end{array}
  \right)
\end{eqnarray}
and the saturation function is:
\begin{equation}
      f(\vect{x}) = c \, \vect{x}^\gamma \,\,\,\,\,\, \textrm{where} \,\,\,\,\,\, \gamma = 0.4, \,\,\,\,\,\,  \textrm{and,}
\end{equation}
the scaling constant is, $c = \hat{\vect{x}}^{1-\gamma}$, and $\hat{\vect{x}}$ is the average response over natural images (for the Divisive Normalization transform):
\begin{equation*}
      \hat{\vect{x}} = \left(
  \begin{array}{c}
    1.12 \\ 
    0.02 \\
    0.01 \\
  \end{array}
\right)
\end{equation*}
This exponent is also used for the definition of energy in Wilson-Cowan, $\vect{e} = |\vect{r}|^\gamma$.
\end{itemize}

Note that the interaction neighborhoods have unit volume, $\sum_j W_{ij} = 1 \,\,\, \forall j$, as suggested in~\cite{Watson97}, and then, the Divisive Normalization kernel is given by the product of this unit-volume neighborhood and two left and right filters in the diagonal matrices, $\mathbb{D}_l$ and $\mathbb{D}_r$~\cite{Martinez19}. The values for the semisaturation, $\vect{b}$, and the diagonal matrices $\mathbb{D}_l$ and $\mathbb{D}_r$ were inspired by the contrast response results in~\cite{Martinez19}: we set the semisaturation according to the average response of natural images (low-pass in nature), and we initialized the left and right filters to high-pass.
However, afterwards, in order to make $\mathcal{N}_{\textrm{DN}}$ and $\mathcal{N}_{\textrm{WC}}$ consistent, we
applied the Divisive Normalization over natural images and we iteratively updated the values of the right and left filters
according to Eq.~\ref{relation_W_H}.
In the end, we arrived to the values in the above expressions (where the filter at the left is high-pass, but the filter at
the right is not).
Note that the attenuation in Wilson-Cowan is computed using Eq.~\ref{relation_W_H}.

\paragraph{Jacobian.} The information theoretic computations strongly depend on how the system (locally) deforms the
signal representation (e.g. Eq.~\ref{analitic}). This is described by the Jacobian of the transform with regard to
the signal, $\nabla_{\vect{r}^1} S = \nabla_{\vect{r}^2} \mathcal{N}^{(2)} \cdot \nabla_{\vect{x}^1} \mathcal{L}^{(2)} \cdot \nabla_{\vect{r}^1} \mathcal{N}^{(1)}$. In this reduced-scale model, this Jacobian (for the Wilson-Cowan case) is:
\begin{equation}
      \nabla_{\vect{r}^1} S = \gamma_2 \left(\mathbb{D}_{\vect{\alpha}} + \vect{W} \cdot \mathbb{D}_{\frac{df}{dx}}\right)^{-1}
\cdot \left( \mathbb{D}_{\left( \mathbb{D}_{\vect{\alpha}} \cdot \vect{x}^2 + \vect{W} \cdot f(\vect{x}^2) \right)}\right)^{1-\frac{1}{\gamma_2}} \cdot G_{\textrm{CSF}} \cdot F \cdot \mathbb{D}_{\gamma_1 (\vect{r}^1)^{\gamma_1-1}}
\end{equation}


\acknowledgments

This work was partially funded by the Spanish Government and EU FEDER fund through the MINECO grants TIN2015-71537-P and DPI2017-89867-C2-2-R; and by the European Union's Horizon 2020 research and innovation programme under grant agreement number 761544 (project HDR4EU) and under grant agreement number 780470 (project SAUCE).


\bibliographystyle{JHEP}
\bibliography{biblio,JesusMaloRef,references_mb,refs_butts}

%
%
%
%
%
%
%
%
\end{document}